\documentclass[preprint,review,12pt]{elsarticle}

\usepackage{amsmath, amssymb, amsfonts, amsthm, amsbsy, mathtools}
\newcommand{\abs}[1]{\ensuremath{\left\vert#1\right\vert}}
\newcommand\scalemath[2]{\scalebox{#1}{\mbox{\ensuremath{\displaystyle #2}}}}

\usepackage{textcomp}

\usepackage{hyperref}

\usepackage{caption}

\usepackage{lineno}
\modulolinenumbers[1]

\journal{Journal of Power Sources}

\biboptions{square, compress}

\begin{document}

\begin{frontmatter}

\title{Characterization of gas diffusion electrodes for metal-air batteries}

\author[DLR,HIU]{Timo Danner\corref{cor1}}
\cortext[cor1]{Corresponding author:}
\ead{Timo.Danner@DLR.de}
\ead[url]{http://www.dlr.de/tt/en/}
\address[DLR]{Institute of Engineering Thermodynamics, German Aerospace Center (DLR), Pfaffenwaldring 38-40, 70569 Stuttgart, Germany}
\address[HIU]{Helmholtz Institute Ulm for Electrochemical Energy Storage (HIU), Helmholtzstra\ss e 11, 89081 Ulm, Germany}
\address[UU2]{Electron Microscopy Group of Materials Science, University of Ulm, Albert-Einstein-Allee 11, 89081 Ulm, Germany}
\address[LIST]{Advanced Instrumentation for Ion Nano-Analytics (AINA/MRT),Luxembourg Institute of Science and Technology, Rue du Brill 41, L-4422 Belvaux, Luxembourg}
\address[DHBW]{Engineering department, Baden-Wuerttemberg Cooperative State University Mannheim, Coblitzallee 1-9, 68163 Mannheim, Germany}
\address[UU1]{Institute of Electrochemistry, University of Ulm, Albert-Einstein-Allee 47, 89081 Ulm, Germany}

\author[UU2,LIST]{Santhana Eswara}

\author[DHBW]{Volker P. Schulz}

\author[DLR,HIU,UU1]{Arnulf Latz}

\begin{abstract}
Gas diffusion electrodes are commonly used in high energy density metal-air batteries for the supply of oxygen. Hydrophobic binder materials ensure the coexistence of gas and liquid phase in the pore network. The phase distribution has a strong influence on transport processes and electrochemical reactions. In this article we present 2D and 3D Rothman-Keller type multiphase Lattice-Boltzmann models which take into account the heterogeneous wetting behavior of gas diffusion electrodes. The simulations are performed on FIB-SEM 3D reconstructions of an Ag model electrode for predefined saturation of the pore space with the liquid phase. The resulting pressure-saturation characteristics and transport correlations are important input parameters for modeling approaches on the continuum scale and allow for an efficient development of improved gas diffusion electrodes.
\end{abstract}

\begin{keyword}
Lattice-Boltzmann method, gas diffusion electrodes, FIB-SEM tomography, metal-air batteries, multiphase flow
\end{keyword}

\end{frontmatter}


\section{Introduction}
\label{sec:introduction}

Metal-air batteries possess a very high theoretical energy density which makes them interesting for both mobile and stationary applications \cite{Rahman:MetalAirReview, Cheng2012:MetalAirReview}. At the negative electrode, metals like Al \cite{Egan:AluminumAirReview}, Li \cite{Girishkumar2010:LiAirPromiseChallenges}, Mg \cite{Zhang2015:MgAir}, Na \cite{Palomares:NaReview, Slater:NaReview}, and Zn \cite{Lee2011:ZnAir, Li2014:ZnAir} were suggested in the literature \cite{Kim:MetalElectrodesReview}. In recent years Li-air batteries received the most attention in the battery community \cite{Christensen2012:LiO2Review, ShaoHorn:BridgingMechanisticUnderstanding}. However, the only system which successfully reached the stage of mass production is the primary Zn-air battery. At the positive electrode oxygen is reduced and evolved during discharge and charge, respectively. Sufficient supply of O$_2$ during discharge is accomplished by the concept of porous gas diffusion electrodes (GDEs). The electrodes are commonly made of carbon materials. However, in aqueous systems carbon is known to dissolve ('carbon corrosion') and carbon-free GDEs were proposed \cite{Wittmaier2014:Ag, Wittmaier2014:Ni, Wittmaier2015:AgCo}. Hydrophobic binder materials ensure the coexistence of gas and liquid phase in the porous structure of the GDE. The saturation behavior is characteristic for the porous material and can be described by capillary pressure saturation ($p_c-s$) curves. The amount and distribution of the liquid phase has a strong influence on transport processes. The transport in the gas phase ensures a good supply of O$_2$ and, thus, allows to draw high current densities. Moreover, the binder improves the mechanical stability of the electrode.\\
Due to their application in alkaline fuel cells the concept of gas diffusion electrodes was studied already in the 1960s. Design optimizations of the electrodes were mainly done by intensive experimental studies. In recent years the improvements in computational efficiency made computer simulations a common design tool in the engineering disciplines. However, traditional CFD (computational fluid dynamics) tools like the volume of fluid (VOF) method \cite{Hirt1981:VOF, Ferreira2015:VOF} have their limitations in the simulation of multiphase flow in complex geometries. In recent years the lattice Boltzmann method (LBM) \cite{McNamara1988:LBM, Sukop2006:LBM, Succi2001:LBM} became increasingly popular for this class of problems because it is easy to implement and scales favorably. In LBM a probability distribution of discrete particle velocities is propagated on a computational lattice. Interactions between particles, boundaries, etc. are modeled by suitable collision operators. Several multiphase models were suggested in the literature for the simulation of immiscible fluids \cite{Sukop2015:MultiphaseLBM}. The most prominent ones are the Shan-Chen model  \cite{Shan1993:SC1, Shan1993:SC2}, the free-energy model \cite{Swift1995:FE1, Swift1996:FE2}, and the color gradient or Rothman-Keller (RK) model \cite{Rothman1988:RK, Gunstensen1991:RK1, Gunstensen1992:RK2}. A common problem of the methods is the numerical stability and accuracy in the simulation of systems with high density and viscosity ratios which also includes the air-water system. Recent publications present modifications of the models which are able to overcome this limitation \cite{Inamuro2004:FE1, Zheng2006}. This extends the applicability of the method to technical systems like gas diffusion media of polymer electrolyte fuel cells and the number of publications in the field increased rapidly \cite{Niu2007:LBM, Park2008:LBM, Mukherjee2011:Review, Hao2012:CapillaryPressure, Nabovati2014:PorosityEffect, Dong2015:LBM_compression}. However, to our knowledge this is the first publication of pore-scale LBM simulations of gas diffusion electrodes for metal-air batteries. In our work we use an RK type multiphase model to simulate multiphase flow in two and three dimensions. The model is based on the work of Leclaire \cite{Leclaire2011:ColorGradient, Leclaire2012:Recoloring, Leclaire2013:MultiPhase} and Liu \textit{et al.} \cite{Liu2012:RK3D}. In their publications the authors successfully demonstrated the simulation of high density ratios. In our study we focus on aqueous electrolytes, however, the presented methodology is not limited to this case. An important feature of our model is that we explicitly take into account the non-homogeneous wetting properties of the GDE which consists of hydrophilic electrode particles and hydrophobic binder fibers. This is an important step and has been barely pursued in previous pore-scale studies of electrochemical devices \cite{Hao2012:CapillaryPressure, Molaeimanesh:PTFE, Hao2010:Permeabilities, Hao2010:WaterTransport, Zhou2010:WaterTransport, Mukherjee2009:WaterManagement}.\\
The focus of our article is set on the development of a methodology for the characterization of gas diffusion electrodes for metal-air batteries. First, the structure of porous Ag model electrodes is reconstructed based on FIB-SEM tomography as explained in Section \ref{sec:reconstruction}. The reconstruction serves as simulation domain for multiphase LBM simulations. Model equations and computational details are summarized in Sections \ref{sec:LBM} and \ref{sec:methodology}. 2D and 3D simulations are performed to simulate the evolution of the phase distribution towards an energetic minimum (Section \ref{sec:results_distribution}). The results are evaluated to obtain characteristic pressure-saturation curves (Section \ref{sec:results_pcsw}) and saturation dependent transport parameters (Section \ref{sec:results_transport}).

\section{Electrode reconstruction}
\label{sec:reconstruction}

\begin{figure}[ht]
  \includegraphics[width=\textwidth,keepaspectratio=true]{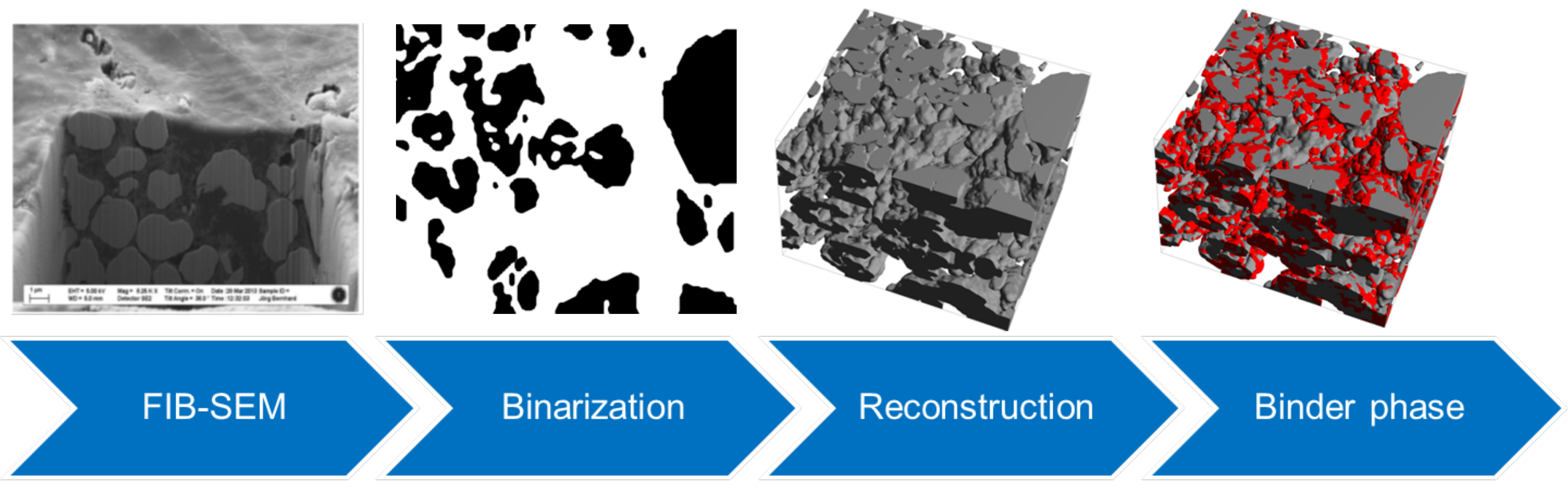}
 \caption{Methodology to reconstruct the LBM simulation domain from FIB-SEM images. Full details regarding the dimensions and voxel sizes are given in Table \ref{tab:recon_StrucData}. Scale bar of SEM image is 1 \textmu m.}
 \label{fig:recon_Reconstruction}
\end{figure}%

We use samples of Ag model electrodes which were characterized regarding their electrochemical performance in our previous article \cite{Danner2014:Halfcell}. The focus of this work is set on the structural characterization and investigation of transport processes on the pore-scale. In this section we explain in detail the methodology which was developed for the reconstruction of gas diffusion electrodes using FIB-SEM tomography. The suggested procedure is schematically shown in Figure \ref{fig:recon_Reconstruction}.

\subsection{FIB-SEM}
\label{sec:recon_FIBSEM}

To obtain the micro-structure for our simulations a dual-beam ZEISS N-Vision 40 SEM-FIB instrument is used. The SEM has a field-emission electron gun with an acceleration voltage between 5 and 30 kV and a vertical electron-optic axis. The FIB is based on a Ga$^+$ primary ion-beam with a 30 kV acceleration voltage. The ion-optic axis is at an angle of 54\textdegree$\,$ to the electron-optic axis. The standard vacuum levels for the electron-gun and the sample chamber are 10$^{-9}$ and 10$^{-5}$ mbar, respectively. The serial sectioning of the sample is achieved using FIB and the images are acquired using the SEM at specified milling intervals.\\
Before the SEM imaging of the sample, it must be ensured that the SEM and FIB images correspond to the same region-of-interest of the sample. This condition is obtained in the following way: Firstly, we set the sample stage to the eucentric tilt position to avoid an offset in sample position with sample tilt. Secondly, the co-incidence of the electron and ion beam has to be ensured. This is established by tilting the sample by 54\textdegree$\,$ with a constant working-distance of about 5 mm and adjusting the Z-axis until the SEM image of the sample comes into focus. A trench is cut into the sample in the vicinity of the region-of-interest with a relatively intense ion-beam current of 6.5 nA such that the cross-sectional (CS) plane becomes visible. The CS plane is then gently polished with an ion-beam current of 300 pA. Subsequently, the total region for 3D tomography is selected for ion-milling. The two samples were filled with a low viscosity epoxy resin in order to improve the contrast of the images. This is an important step to facilitate the reconstruction process. The left panel of Figure \ref{fig:recon_Reconstruction} shows a representative SEM image. We note that the resin has rather well impregnated the pore space of the GDEs. Horizontal and vertical dimension are in the following named $x$ and $y$, respectively. The direction perpendicular to the $x$-$y$ plane is denoted by $z$ and represents the direction of the FIB cut. The total thickness of the cut is 10 \textmu m. In order to optimize the total duration of milling and image-acquisition, the SEM images are obtained at every fifth FIB-slice which results in 84 images representing slices of 0.12 \textmu m thickness. Note, that ($x$, $y$) values of the voxel-size are related by the relation $y$ = $x$/sin(54\textdegree), where the value of $x$ is determined by the magnification of the image and the 54\textdegree$\,$ origin from the tilt of the sample.

\subsection{Structure generation}
\label{sec:recon_structure}

The second panel of Figure \ref{fig:recon_Reconstruction} shows a representative binarized and cropped image of the electrode micro-structure (black). The images of different slices were aligned to account for the sample-drift during imaging. The perspective correction and pixel-size adjustments were done in the software package IMOD. The epoxy resin improves the contrast in the images and helps to identify solid particles. In spots where the impregnation of the electrode is incomplete the phases are assigned manually. Further details of the methodology of reconstruction is discussed in detail elsewhere \cite{Eswara2014:FIB-SEM}. Finally, the images were stacked to a virtual structure in the commercial software GeoDict \cite{GeoDict:UserGuide}. The resulting geometry can be seen in the third panel of Figure \ref{fig:recon_Reconstruction}.\\
The reconstructions are mirrored at the $x-y$ plane in order to increase the simulation domain in the direction of FIB sampling ($z$-direction). This step avoids systematic errors due to periodic boundary conditions which are used in the LBM simulations. The resulting geometry is subsequently coarsened in order to decrease the computational load. The dimensions of the samples are summarized in Table \ref{tab:recon_StrucData}.\\
As outlined above the hydrophobic binder material is important for the distribution of the liquid phase and, thus, performance of the device. Unfortunately, the binder distribution is lost in the imaging and reconstruction process. In our reconstructions we distribute binder fibers by a random walk algorithm on the electrode surface. Binder fibers crossing the void space between electrode particles are neglected. The resulting binder distribution is shown in red color in the right panel of Figure \ref{fig:recon_Reconstruction}. It is in reasonable optical agreement with distributions observed in SEM images \cite{Danner2014:Halfcell}.

\section{Lattice Boltzmann Method}
\label{sec:LBM}

\subsection{Model description}
\label{sec:LBM_model}

In this work we perform two (D2Q9) and three (D3Q19) dimensional simulations of two-phase flow in porous gas diffusion electrodes. The LBM multiphase models employed in this study are based on the color-gradient approach and follow the recent publications of Liu and Leclaire \textit{et al.} \cite{Leclaire2011:ColorGradient, Leclaire2012:Recoloring, Leclaire2013:MultiPhase, Liu2012:RK3D}. A detailed derivation and discussion of the model equations can be found in their publications.\\
In RK type models the state of each phase $k$ ($k$=gas, liquid) is described with the help of a probability distribution function $f^k_i(\mathbf{x}, t)$, where $i$ is the index representing discrete directions in velocity space (see Eq. \eqref{eq:appendix_D2Q9Lattice} and \eqref{eq:appendix_D3Q19Lattice}). The macroscopic properties at a lattice node $\mathbf{x}$ are given by the moments of the probability distribution. For instance, the density $\rho$ is calculated by the 0th moment which is simply the sum of the probabilities in all lattice directions $nv$
\begin{equation}
 \rho = \sum_i^{nv} f_i\;.
 \label{eq:methodology_MacroscopicDensity}
\end{equation}
The fluid velocity $\mathbf{v}$ follows as the 1st moment which can be regarded as average of the discrete velocities $\mathbf{e}$ weighted by their probability
\begin{equation}
 \mathbf{v} =  \frac{1}{\rho}\sum_i^{nv} \mathbf{e}_i f_i\;.
 \label{eq:methodology_MacroscopicVelocity}
\end{equation}
Finally, the pressure can be calculated according to
\begin{eqnarray}
p_k &=& \rho_k(c_s^k)^2 = \frac{3}{5}\rho_k(1-\alpha_k) \quad  \text{(D2Q9)} \nonumber\\
p_k &=& \rho_k(c_s^k)^2 = \frac{1}{2}\rho_k(1-\alpha_k) \quad  \text{(D3Q19)}\;,
 \label{eq:methodology_PressureLattice}
\end{eqnarray}
where $\alpha_k$ is a simulation parameter given in \ref{sec:appendix_parameters}.\\
In general the temporal evolution of the probability distribution $f^k_i(\mathbf{x}, t)$ is described by
\begin{equation}
  f^k_i(\mathbf{x}+\mathbf{e}_i, t+\Delta t) = f^k_i(\mathbf{x}, t) + \Omega^k_i \left(f^k_i(\mathbf{x}, t)\right)\;,
  \label{eq:methodology_LBMalghorithm}
\end{equation}
where the collision operator $\Omega_i^k$ is in RK type models a result of three sub operators \cite{Tolke2002:RK, Reis2007:RK}
\begin{equation}
  \Omega^k_i = \Omega^{k,1}_i \left(\Omega^{k,3}_i+\Omega^{k,2}_i\right)\;.
  \label{eq:methodology_CollisionOperator}
\end{equation}

The operators are applied successively to the probability distribution $f^k_i(\mathbf{x}, t)$ and are defined by
\begin{enumerate}
  \item Single-phase collision (SRT-BGK)
  \begin{equation}
    f^k_i(\mathbf{x}, t^*) = f^k_i(\mathbf{x}, t) + \Omega^{k,1}_i \left(f^k_i(\mathbf{x}, t)\right)\;
    \label{eq:methodology_CollisionBGK}
  \end{equation}
  In our model we apply the standard SRT-BGK approximation \cite{Bhatnagar1954:BGKOperator} to model fluid interactions in the same phase
  \begin{equation}
  \Omega_i^{k,1} = -\frac{f_i^k(\mathbf{x},t)-f_i^{k, \text{eq}}(\mathbf{x},t)}{\tau_k} \;,
  \label{eq:methodology_CollisionOperatorBGK}
  \end{equation}
  where $\tau_k$ is the relaxation time of the collision process and $f_i^{k, \text{eq}}$ the local equilibrium distribution of fluid velocities. The relaxation time $\tau_k$ can be related to the kinematic viscosity $\nu_k$ by
  \begin{equation}
  \tau_k =  \frac{3\nu_k}{c^2}+0.5\Delta t\;.
  \label{eq:methodology_RelaxationTime}
  \end{equation}
  In our simulations we use a single relaxation time $\tau_\text{gas}=\tau_\text{liq}=\bar{\tau}$ for both phases which is calculated from a density average of the kinematic viscosity 
  \begin{equation}
    \bar{\nu} = \left(\sum_k \frac{\rho_k}{\rho \nu_k}\right)^{-1}\;.
    \label{eq:methodology_ViscosityAverage}
  \end{equation}
  The local equilibrium distribution $f^{k, \text{eq}}$ is given by the Maxwell distributions
  \begin{equation}
  f_i^{k, \text{eq}} = \rho \left( \phi^k_i + w_i\left[ 3\frac{\mathbf{e}_i\mathbf{v}}{c^2} + \frac{9}{2}\frac{(\mathbf{e}_i\mathbf{v})^2}{c^4} - \frac{3}{2}\frac{\mathbf{v}^2}{c^2}\right] \right)\;,
  \label{eq:methodology_MaxwellEquilibrium}
  \end{equation}
  where $\phi^k_i$ (Eq. \eqref{eq:appendix_phiD2Q9} and \eqref{eq:appendix_phiD3Q19}) determines the compressibility of the fluid and $w_i$ is a lattice specific weighting parameter (see Eq. \eqref{eq:appendix_wD2Q9} and \eqref{eq:appendix_wD3Q19}).

  \item Two-phase collision (Perturbation)
  
  \begin{equation}
    f^k_i(\mathbf{x}, t^{**}) = f^k_i(\mathbf{x}, t^*) + \Omega^{k,2}_i \left(f^k_i(\mathbf{x}, t^*)\right)\;
    \label{eq:methodology_CollisionPerturb}
  \end{equation}
  The effect of surface tension between the two phases is modeled by the perturbation operator. In order to distinguish the phases it is convenient to introduce a color-field $\psi(\mathbf{x},t)$
  \begin{equation}
    \psi(\mathbf{x},t) = \frac{\rho_\text{gas}-\rho_\text{liq}}{\rho_\text{gas}+\rho_\text{liq}}\;.
    \label{eq:methodology_Psi}
  \end{equation}
  Regions of gas and liquid phase are marked by values of $\psi$ close to 1 and -1, respectively. The perturbation operator takes the form \cite{Reis2007:RK, Liu2012:RK3D}
  \begin{equation}
  \Omega^{k,2}_i \left(f^k_i(\mathbf{x}, t^*)\right) = \frac{A_k}{2}\abs{\triangledown \psi}\left[ w_i \frac{(\mathbf{e}_i\triangledown \psi)^2}{\abs{\triangledown \psi}^2}-B_i \right]\,.
  \label{eq:methodology_CollisionPerturbationOperator}
  \end{equation}
  $A_k$ is a parameter controlling the surface tension (see Eq. \eqref{eq:appendix_SurfaceTension}), $\triangledown \psi$ is the color gradient in the two-phase region, and $B_i$ is a parameter which has to be chosen in order to recover the Navier-Stokes Equations \cite{Reis2007:RK, Liu2012:RK3D} (see Eq. \eqref{eq:appendix_BD2Q9} and \eqref{eq:appendix_BD3Q19}). The color gradient determining the surface normal in the two-phase region is approximated by higher-order isotropic discretization schemes which significantly reduce spurious velocities \cite{Sbragaglia2007, Liu2012:RK3D, Leclaire2011:ColorGradient}.
  
  \item Two-phase collision (Recoloring)
  \begin{equation}
    f^k_i(\mathbf{x}, t^{***}) = \Omega^{k,3}_i \left(f^k_i(\mathbf{x}, t^{**})\right)\;
    \label{eq:methodology_CollisionRecoloring}
  \end{equation}
  The perturbation operator models the effect of surface tension, however, it does not enforce phase separation. This is ensured by the recoloring operator
  \begin{eqnarray}
    \Omega^{\text{gas},3}_i\left(f^\text{gas}_i(\mathbf{x}, t^{**})\right)\ = \frac{\rho_\text{gas}}{\rho}f_i  +  \beta \frac{\rho_\text{gas} \rho_\text{liq}}{\rho^2}\text{cos}\left(\phi_i \right) f_i^\text{eq}\vert_{\mathbf{v}=0} \; \nonumber\\
    \Omega^{\text{liq},3}_i\left(f^\text{liq}_i(\mathbf{x}, t^{**})\right)\ = \frac{\rho_\text{liq}}{\rho}f_i  -  \beta \frac{\rho_\text{gas} \rho_\text{liq}}{\rho^2}\text{cos}\left(\phi_i \right) f_i^\text{eq}\vert_{\mathbf{v}=0} \;,
    \label{eq:methodology_CollisionRecoloringOperator}
  \end{eqnarray}
  where $\rho=\rho_\text{gas}+\rho_\text{liq}$ is the total density, $f_i=f_i^\text{gas}+f_i^\text{liq}$ the total probability distribution, and $\phi_i$ the angle between the color gradient and the lattice direction vector 
  \begin{equation}
    \text{cos}\left(\phi_i \right) = \frac{\mathbf{e}_i \triangledown \psi}{\abs{\mathbf{e}_i}\abs{\triangledown \psi}}\;.
    \label{eq:methodology_CollisionRecoloringPhi}
  \end{equation}
  The operator allows a moderate mixing of the two phases at the interface which additionally reduces spurious currents. The interface thickness is controlled by the parameter $\beta$. Moreover, it was shown that the proposed operator circumvents the problem of 'lattice-pinning' \cite{Latva-Kokko2005:LatticePinning}.

  \item Streaming
  \begin{equation}
    f^k_i(\mathbf{x}+\mathbf{e}_i, t+1) = f^k_i(\mathbf{x}, t^{***})\;
    \label{eq:methodology_Streaming}
  \end{equation}
  Finally, the modified probability distributions are streamed to their new positions on the computational grid and the corresponding boundary conditions are applied (see Section \ref{sec:LBM_boundary}).
\end{enumerate}

\subsection{Boundary conditions}
\label{sec:LBM_boundary}
There are two types of boundaries in micro-structure resolved simulations of porous media: boundaries of the computational domain and boundaries at the fluid-solid interface. For the first type we employ simple periodic boundary conditions. This choice has implications on the simulation procedure and will be discussed in Section \ref{sec:methodology_initial}.\\
At the fluid-solid interface we use a simple bounce back scheme. The contact angle is adjusted by assigning an effective density to the probability distribution at solid nodes \cite{Latva-Kokko2005:ContactAngle}. This has an effect on the local color-gradient and determines if the surface is wetting or non-wetting. In our model the densities on solid nodes were adjusted to match the contact angles which are observed experimentally (see Figure \ref{fig:LBM_ModelValidation}). Parameters can be found in Table \ref{tab:LBM_Parameters}.

\subsection{Model parameterization}
\label{sec:LBM_parameterization}
In this study we parameterize our models for the air-water system in order to represent metal-air batteries with aqueous electrolytes. It has to be noted that other electrolyte systems using organic solvents or ionic liquids can be treated in the same framework. An analysis of dimensionless numbers is helpful to evaluate the relevant forces of the physical problem. Fluid flow in porous media is mainly determined by gravitational, viscous, inertial, and surface forces. The dimensionless numbers describing the ratio between these forces are
\begin{itemize}
 \item Bond number
\begin{equation}
    Bo = \frac{\text{gravitational forces}}{\text{surface forces}} = \frac{\rho \; g \; l^2}{\sigma} \sim 4 \cdot 10^{-7} 
 \label{eq:methodology_BondNumber}
\end{equation}
 \item Capillary number
\begin{equation}
    Ca = \frac{\text{viscous forces}}{\text{surface forces}} = \frac{\mu \; \mathbf{v}}{\sigma} \sim 1 \cdot 10^{-8} 
 \label{eq:methodology_CapillaryNumber}
\end{equation}
 \item Reynolds number
\begin{equation}
    Re = \frac{\text{inertial forces}}{\text{viscous forces}} = \frac{\rho \; \mathbf{v} \; l}{\mu} \sim 2 \cdot 10^{-6}
 \label{eq:methodology_ReynoldsNumber}
\end{equation}
\end{itemize}
The characteristic length $l$ is in our studies given by the mean pore diameter $d_{50}=0.87$ \textmu m of the electrode. The flow velocity $\mathbf{v}$ can be approximated by the flow rate of volume replacement experiments and is typically less than the $1 \cdot 10^{-6}$ m s$^{-1}$ used in above calculations \cite{Dwenger2011:CapillaryPressure}. Moreover, in the limit of fluid mechanical equilibrium it will tend to zero. The small values of the dimensionless numbers demonstrate that gravitational and viscous forces are negligible compared to the strong influence of surface forces. Therefore, we use unity density and viscosity ratios in our simulations to improve the numerical stability. In a post processing step suitable scaling rules are applied to the simulation results \cite{Schaap2007:Rescaling}. The Laplace equation
\begin{equation}
    \Delta p_c = \frac{\sigma}{r}\;\text{(2D)}\;\; \text{and}\;\; \Delta p_c = \frac{2\sigma}{r}\;\text{(3D)}
 \label{eq:LBM_Laplace}
\end{equation}
allows to deduce a relationship between the capillary pressure of the 'physical' ($\Delta p_c$) and simulated system ($\Delta p^\text{LBM}_c$) according to
\begin{equation}
    \Delta p_c = \frac{\sigma}{\sigma^\text{LBM}}\frac{1\;(lu)}{\Delta x}\Delta p^\text{LBM}_c\;,
 \label{eq:methodology_PressureScaling}
\end{equation}
where $lu$ is one lattice unit, $\Delta x$ the size of one voxel and $\sigma$ and $\sigma ^\text{LBM}$ the surface tension of the physical system and LBM simulation, respectively.
The parameters of this study are summarized in Table \ref{tab:LBM_Parameters}.

\subsection{Model validation}
\label{sec:LBM_validation}

\begin{figure}[ht]
 \includegraphics[width=\textwidth,keepaspectratio=true]{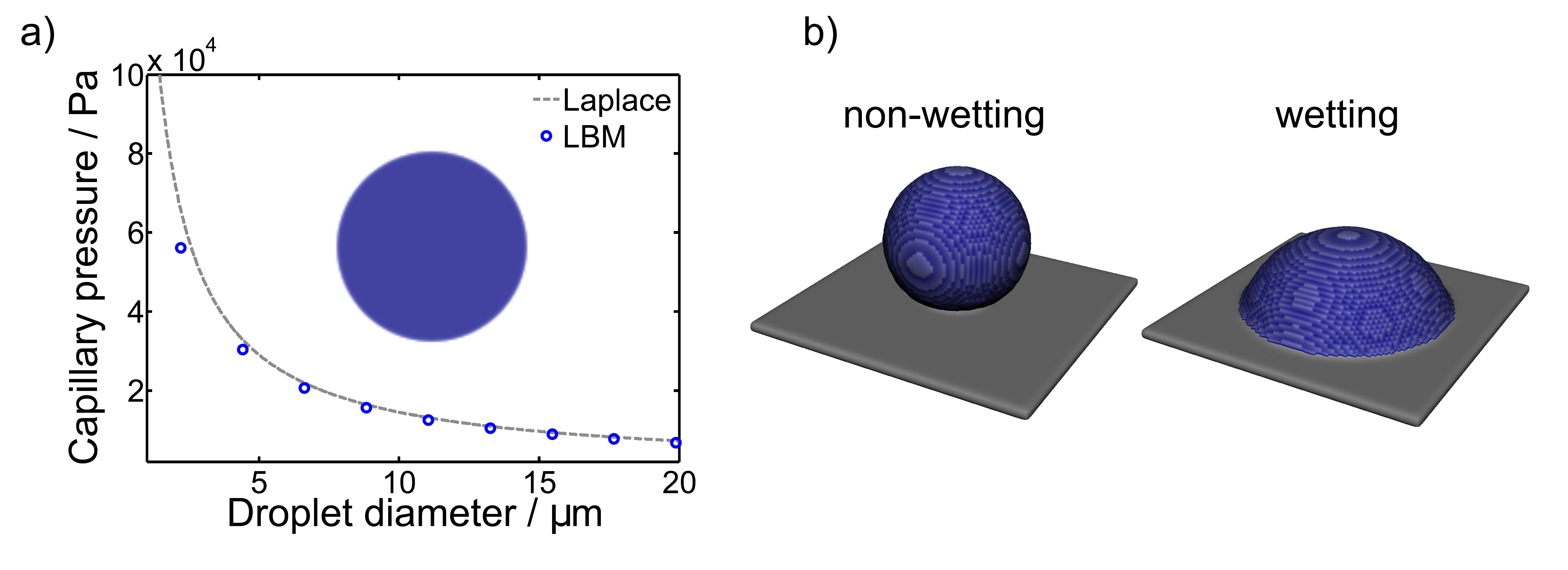}
 \caption{Validation simulations of the LBM model. Left: Bubble test. Right: Influence of non-wetting and wetting surfaces.}
 \label{fig:LBM_ModelValidation}
\end{figure}%

In order to validate our model we present two simple test cases.

\paragraph{Bubble test}
The pressure difference across the interface of a steady bubble can be calculated by the Laplace equation (Eq. \eqref{eq:LBM_Laplace}). In order to validate our model we perform 2D simulations on a lattice with 100x100 nodes and varying bubble diameter. Figure \ref{fig:LBM_ModelValidation} a) shows simulation results for diameters ranging from 10 to 90 lattice units. The size of one lattice unit corresponds to the voxel size of the reconstruction (see Table \ref{tab:recon_StrucData}). The simulated pressures reproduce favorably the Laplace equation. At small diameters a minor deviation of simulation results is observed. This can be assigned to the increasing ratio of surface to bulk voxels which is important to notice for the calculation of $p_c-s$ curves.

\paragraph{Static contact angle}
The second case is the simulation of static contact angles as presented in Figure \ref{fig:LBM_ModelValidation} b). The fluid density of solid nodes was adjusted to reproduce the contact angle of water on Ag \cite{Zhao:ContactAngleAgPTFE} and PTFE surfaces \cite{Mukherjee2009:WaterManagement}. The Figure demonstrates that our model is able to simulate the wetting characteristics of the Ag GDE sample. 

\section{Simulation methodology}
\label{sec:methodology}

\subsection{Initial conditions}
\label{sec:methodology_initial}
The $p_c-s$ curves of porous media are commonly determined with the method of standard porosimetry (MSP) \cite{Gostick2006:CapillaryPressure, Kumbur2007:LeverettIII} or dynamic volume replacement experiments \cite{Fairweather2007:CapillaryPressure, Gostick2008:CapillaryPressure, Nguyen2008:CapillaryPressure, Dwenger2011:CapillaryPressure}. After filling the porous medium to a defined saturation level the corresponding average capillary pressure follows as difference between the pressure in the gas and the liquid phase. It was found that the direction of the process (injection/imbibition or removal/drainage of the liquid phase) causes a hysteresis in the resulting $p_c-s$ curves. It is therefore expected that a different protocol for loading the porous medium with liquid or gas phase will lead to slightly different $p_c-s$ curves. Since the conditions for the standard dynamic volume experiments do not match the conditions in the GDE under dynamic operations, we chose a different simulation set up to determine $p_c-s$ curves. This setup corresponds to establishing a force balance in the volume of the porous medium after initializing the simulation with a maximally wetting or non-wetting condition respectively for a given volume fraction of liquid in the GDE. We argue, that the thus obtained pressure saturation curves are closer to the real operating condition of a GDE in a metal-air cell, where electrolyte is pushed out of the GDE due to the occurrence of solid reaction products in the bulk of the porous medium and not due to the application of external pressure forces \cite{Danner2014:Halfcell}. The different initial configurations within our LBM model with periodic boundary conditions are shown in the first row of Figure \ref{fig:results_Distributions3D}. During drainage the electrolyte will remain preferentially on the hydrophilic electrode particles. We maximize the interface between liquid phase and hydrophilic Ag particles by randomly placing liquid droplets on the electrode surface or, if all surface sites are occupied, in contact with another liquid droplet (configuration I). In the second configuration the liquid phase is introduced in the pore space as one solid block (Figure \ref{fig:results_Distributions3D} right column - configuration II) in order to minimize the interfacial area between gas and liquid phase. This situation is comparable to the imbibition process in the experimental setup where the electrolyte is forced into the porous structure and also occupies areas which are energetically not favorable.  Our simulations are able to reproduce the hysteretic behavior which is also observed in the forced drainage-imbibition experiments. These results will be discussed in more detail in Section \ref{sec:results_pcsw}.

\subsection{Pressure-saturation curves}
\label{sec:methodology_pcs}
For the determination of $p_c-s$ curves we perform independent 2D and 3D simulations at various saturation levels of the GDE sample. The number of iterations was chosen sufficiently large to achieve fluid mechanical equilibrium. At the end of the simulation the capillary pressure follows as $\Delta p_c^\text{LBM} = \bar{p}^\text{liq} - \bar{p}^\text{gas}$, where $\bar{p}^k$ is averaged over the simulation domain. In the case of 2D simulations we conduct at each saturation 10 independent simulations on randomly chosen two-dimensional slices of the reconstruction in order to get statistically significant results.\\
Pressure saturation curves are commonly described with the so called Leverett function \cite{Leverett1940:J, Udell1985:J}
\begin{equation}
    J(s) = \frac{\Delta p_c}{\sigma\abs{cos(\Theta)}}\sqrt{\frac{B_0}{\varepsilon_0}}\;,
 \label{eq:methodology_LeverettStandard}
\end{equation}
where $B_0$ is the permeability and $\varepsilon_0$ the porosity of the electrode. In this study we use a modification of the standard formulation according to Hao \textit{et al.} \cite{Hao2012:CapillaryPressure} to account for the heterogeneous wetting properties of the GDE
\begin{equation}
    J(s) = \frac{\Delta p_c}{\sigma}\sqrt{\frac{B_0}{\varepsilon_0}}\;.
 \label{eq:methodology_LeverettHetero}
\end{equation}

\subsection{Transport parameters}
\label{sec:methodology_transport}
The final phase distribution at the end of the 3D simulations is used as input for the calculation of saturation-dependent effective transport parameters and surface areas in the software GeoDict \cite{GeoDict:UserGuide}. The final fluid distribution is basically regarded as 'frozen' in the porous structure. For the determination of effective transport parameters in the gas phase the voxels of the liquid phase are then treated as solids and vice versa. In their study Garcia \textit{et al.} \cite{Garcia2015:Diffusivity} took a similar approach to determine the diffusivity of partially saturated GDLs. However, they obtained the distribution of the liquid phase from tomographic images instead of two-phase simulations.\\
In the modeling of electrochemical devices the Bruggeman correlation is often used for the approximation of effective transport parameters. The diffusivity which is the ratio between effective and bulk diffusion coefficients is in this approach described by  
\begin{equation}
 D_k^\text{eff}/D_k^0 = \left(\varepsilon_k\right)^\beta \;,
 \label{eq:methodology_Bruggeman}
\end{equation}
where $\varepsilon_k$ is the volume fraction of the transporting phase and $\beta$ the so-called Bruggeman coefficient. The volume fraction of the liquid and gas phase are calculated according to $\varepsilon_\text{liq}=\varepsilon_0 s$ and $\varepsilon_\text{gas}=\varepsilon_0 (1-s)$, respectively.

\begin{figure}[ht!]
  \includegraphics[width=0.9\textwidth,keepaspectratio=true]{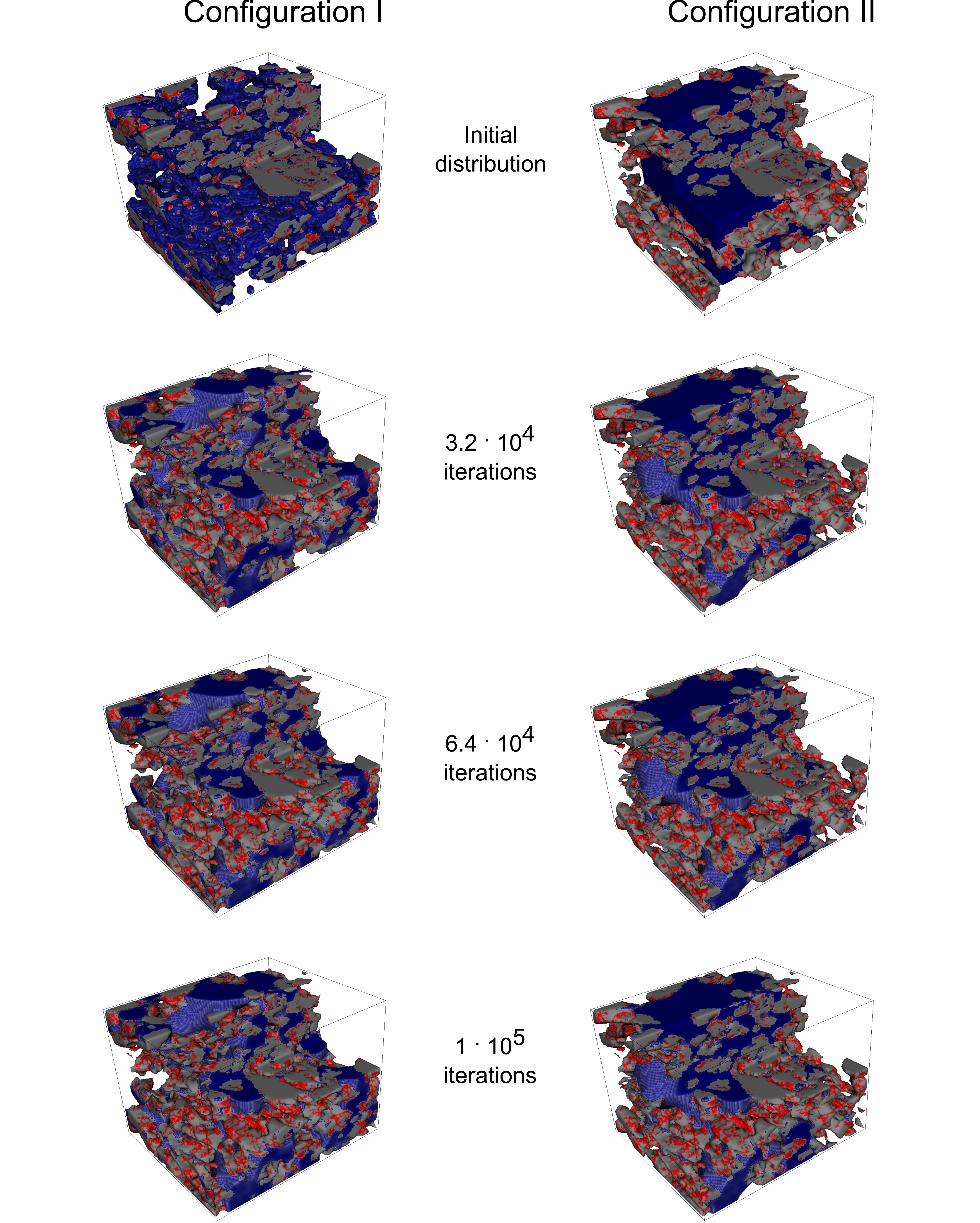}
 \caption{Electrolyte distribution in the porous GDE for configuration I (left) and configuration II (right).}
 \label{fig:results_Distributions3D}
\end{figure}%

\section{Results and discussion}
\label{sec:results}

\subsection{Electrolyte distribution}
\label{sec:results_distribution}

\begin{figure}[ht]
 \includegraphics[width=\textwidth,keepaspectratio=true]{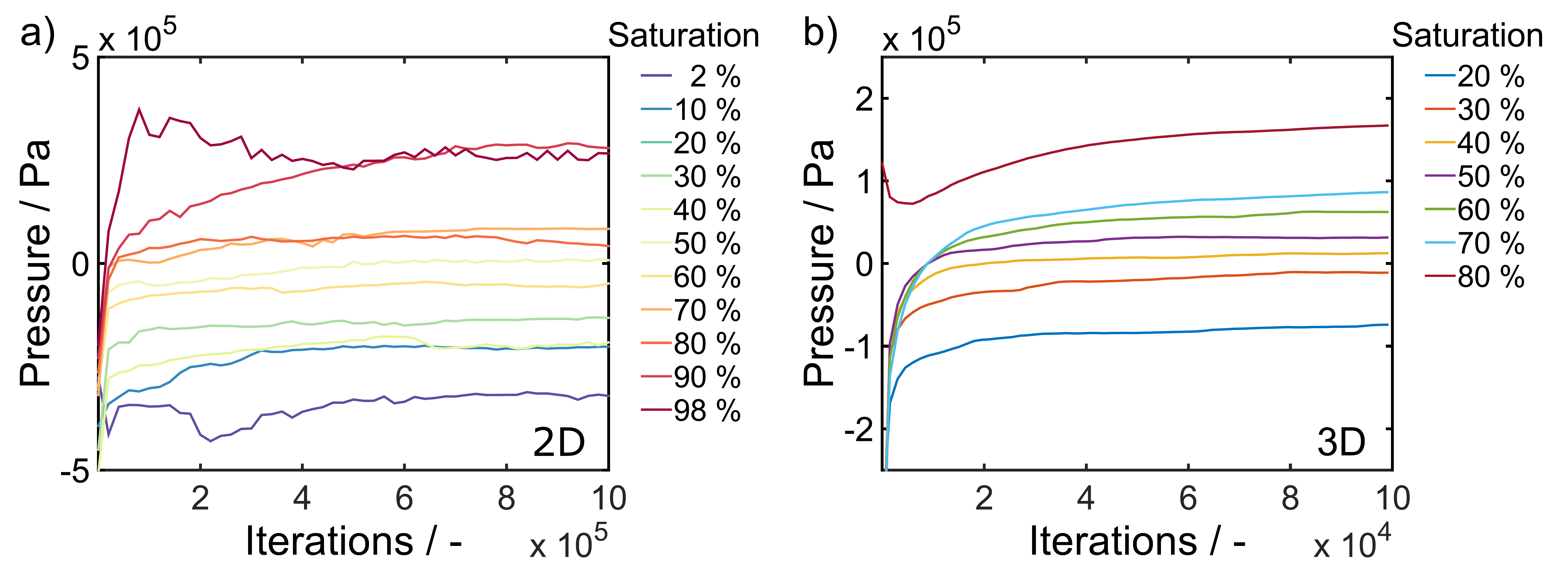}
 \caption{Pressure as function of iterations during simulations employing configuration I. Left: 2D simulations. Right: 3D simulations.}
 \label{fig:results_PressureIterations}
\end{figure}

Figure \ref{fig:results_Distributions3D} shows the phase distribution during 3D simulations of configuration I (left) and configuration II (right) at a saturation of the pore space with liquid electrolyte of 50\%. The first row illustrates the initial conditions which were applied to mimic the drainage and imbibition of the liquid phase. (cf. Section \ref{sec:methodology_initial}). In configuration I the electrolyte is initially distributed on the electrode surface. During the simulation the distribution evolves towards a local minimum of the free energy. Our simulations reach a quasi stationary state after about 1$\cdot$10$^{4}$ iterations. Further changes are only marginal. This is also reflected in the corresponding pressure signal which is shown as function of iterations in Figure \ref{fig:results_PressureIterations} b). A visually similar steady-state is also observed for configuration II. This indicates that the initial condition which we apply in our simulations has only a minor influence on the saturation dependence of effective transport parameters. This effect will be discussed in Section \ref{sec:results_transport}.

\subsection{Pressure-saturation curves}
\label{sec:results_pcsw}

\begin{figure}[t]
 \includegraphics[width=\textwidth,keepaspectratio=true]{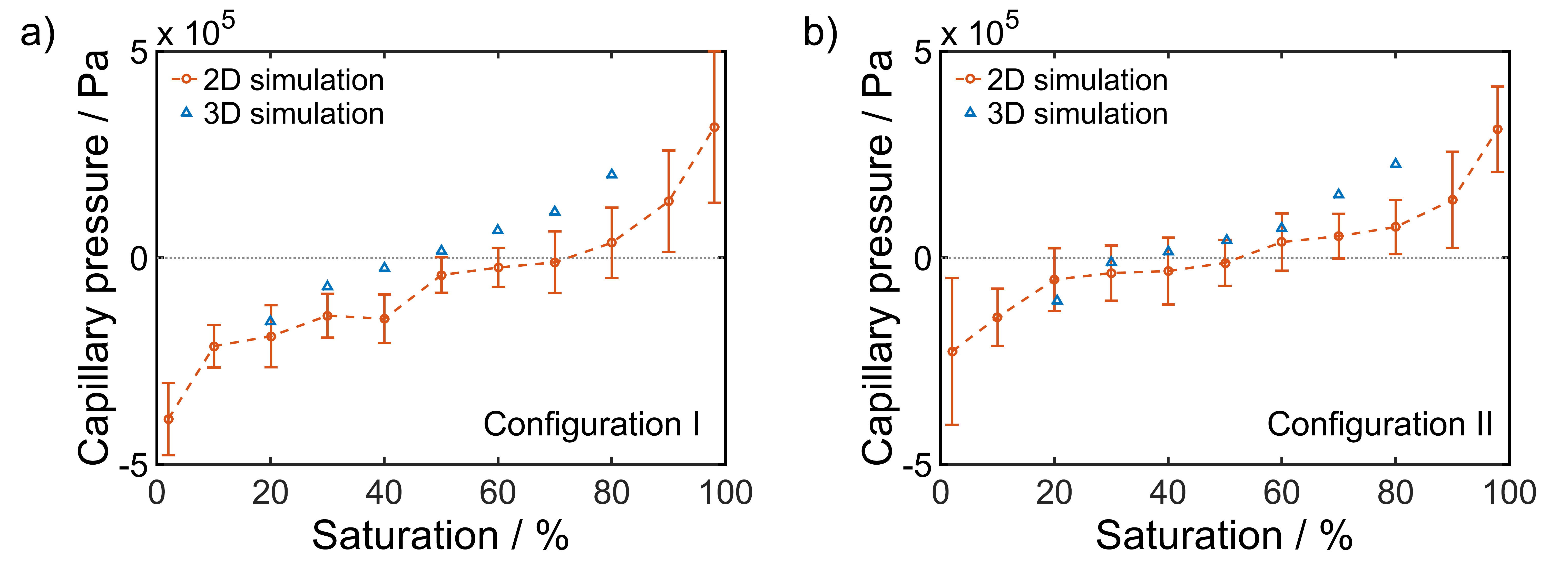}
 \caption{Pressure-saturation curves during for configuration I (left) and configuration II (right). Results of the 2D simulations are shown in red color and error bars represent the standard deviation of the simulations. Results of the 3D simulations are represented by blue triangles.}
 \label{fig:results_PcSwCurves}
\end{figure}%

\begin{figure}[t]
\begin{center}
 \includegraphics[width=0.5\textwidth,keepaspectratio=true]{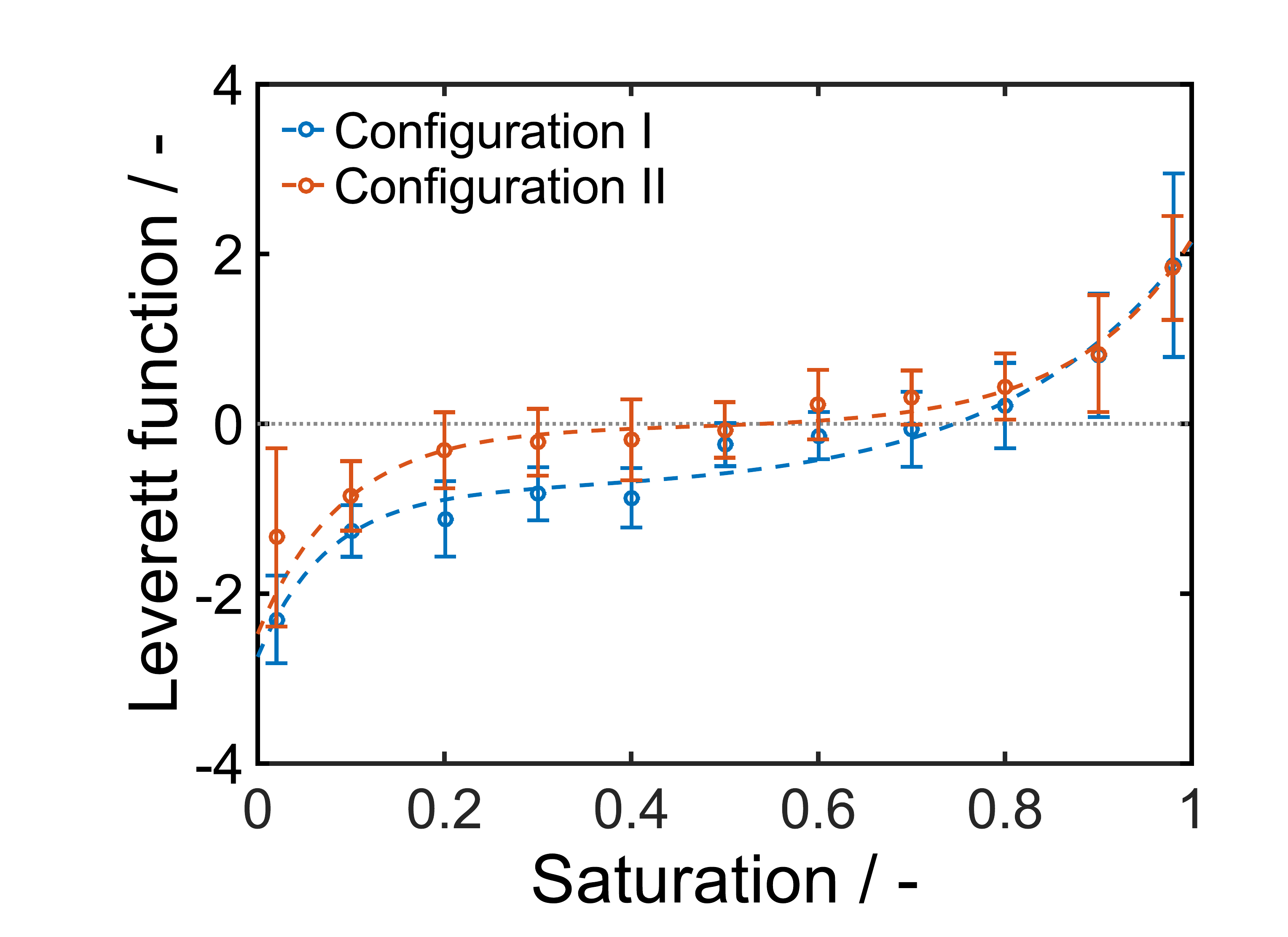}
 \caption{Plot of the dimensionless Leverett Function over saturation for configuration I (blue) and configuration II (red). Dashed lines are a result of the fit to Eq. \eqref{eq:results_LeverettFit}. Symbols represent corresponding simulation results.}
 \label{fig:results_LeverettFunctions}
 \end{center}
\end{figure}

Figure \ref{fig:results_PressureIterations} shows the capillary pressure signal during 2D and 3D simulations representing the drainage process. In both cases the pressure signal approaches a constant value at the end of the simulations which indicates that the system is in a local minimum of the free energy. The 3D simulations give a clear trend of increasing capillary pressure with increasing liquid phase saturation. It has to be noted that the 2D simulations shown in Figure \ref{fig:results_PressureIterations} a) were not performed on the same slice of the reconstruction. Therefore, the fluctuations in pressure which occur are due to different simulation domains and a clear trend can not be deduced from the graph. In our approach we take the average of 10 simulations to capture this effect statistically. Figure \ref{fig:results_PcSwCurves} shows $p_c-s$ curves for configuration I (left) and configuration II (right). The results of the 2D simulations are shown in red color and the error bars represent the corresponding standard deviation. The error bars are larger at low and high saturation which can be explained by a reduced configurational freedom. Meaning, reconfigurations of the phases are suppressed and they remain in their initial (random) configuration. The calculated pressures of 3D simulations are included as blue triangles. There is a systematic deviation at high saturation, however, the general agreement between 2D and 3D simulations is favorable. This is an important result which shows that a series of computationally efficient 2D simulations can be used for a screening of different electrode structures. In a previous publication \cite{Danner2014:Halfcell} we predicted a liquid phase saturation of $\approx$ 50 \%  at $\Delta p=0$ by a fit of continuum simulations to electrochemical data. This is in excellent agreement with the results presented in this work and indicates the validity of our approach.\\ 
Pressure saturation curves are commonly presented as dimensionless Leverett functions (see Eq. \eqref{eq:methodology_LeverettHetero}). Hao \textit{et al.} showed that an expression of the form
\begin{equation}
    J(s) = A+Be^{C(s-0.5)}-De^{-E(s-0.5)}\;
 \label{eq:results_LeverettFit}
\end{equation}
is able to give a good representation of $p_c-s$ data. We fit the parameters of Eq. \eqref{eq:results_LeverettFit} to the $p_c-s$ curves of our 2D simulations. Figure \ref{fig:results_LeverettFunctions} shows Leverett functions of configuration I (left) and configuration II (right). Similar to the experimental work on GDLs we observe a hysteresis at intermediate saturation \cite{Fairweather2007:CapillaryPressure, Dwenger2011:CapillaryPressure}. This can be explained by looking at the experimental procedure and taking into account micro-structural effects in the GDE. Some areas are connected to the pore network only through narrow pores. According to the Laplace equation they are filled with the liquid phase only at high capillary pressures. On the reverse process a higher negative pressure difference is needed to withdraw the electrolyte from this part of the pore network. A similar reasoning can be made for the simulation approach presented in this study. In the simulations starting with configuration I all parts of the pore network are accessible for the liquid phase. This results in a lower capillary pressure compared to simulations with initial configuration II where the access is restricted to surrounding pores. This result justifies our choice of initial conditions to mimic the two processes. However, it has to be noted that the hysteresis is within the standard deviation of the simulations. Experimental studies on the electrodes will be needed to validate our simulation methodology. The parameters of Eq. \eqref{eq:results_LeverettFit} resulting from the fit to the simulation data are summarized in Table \ref{tab:results_LeverettParameters}.

\subsection{Effective transport parameters}
\label{sec:results_transport}

\begin{figure}[ht]
 \includegraphics[width=\textwidth,keepaspectratio=true]{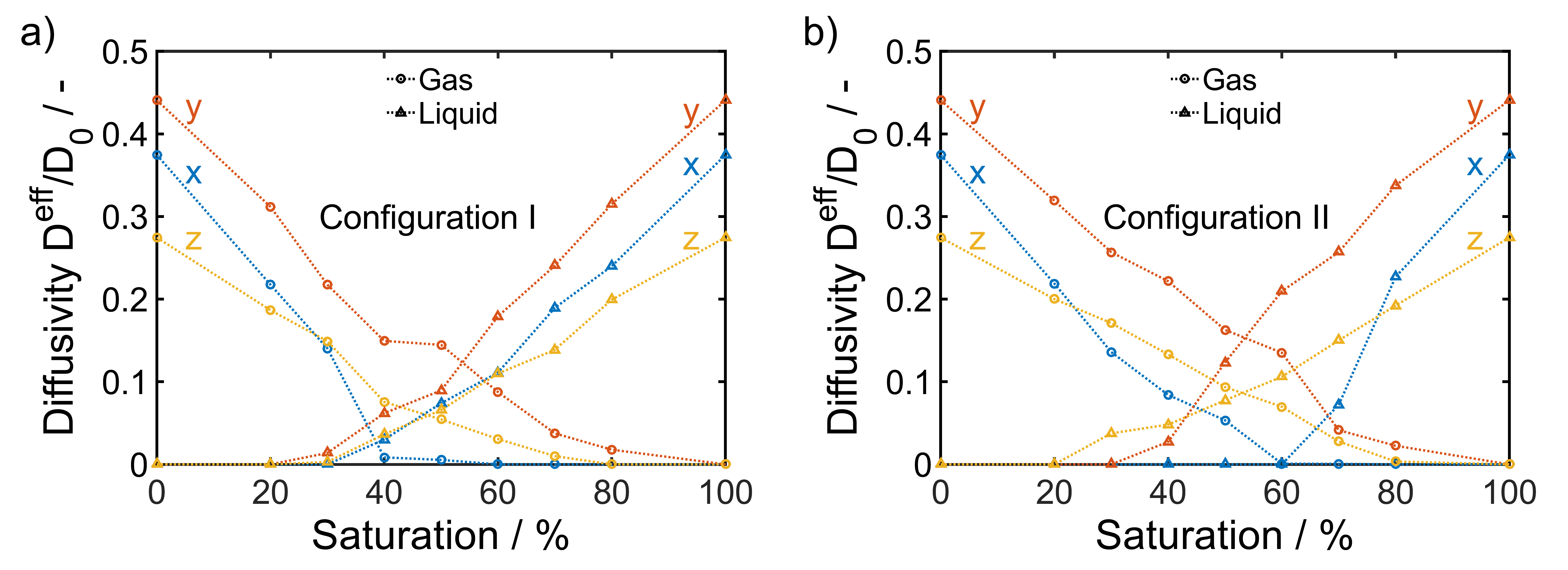}
 \caption{Diffusivity in the gas (open circles) and liquid (open triangles) phase. Values are calculated based on the final electrolyte distribution obtained in the 3D LBM simulations. Plots for the three spatial coordinates ($x$,$y$,$z$) are shown separately for configuration I (left) and configuration II (right).}
 \label{fig:results_Diffusivity}
\end{figure}%

\begin{figure}[t]
\begin{center}
 \includegraphics[width=\textwidth,keepaspectratio=true]{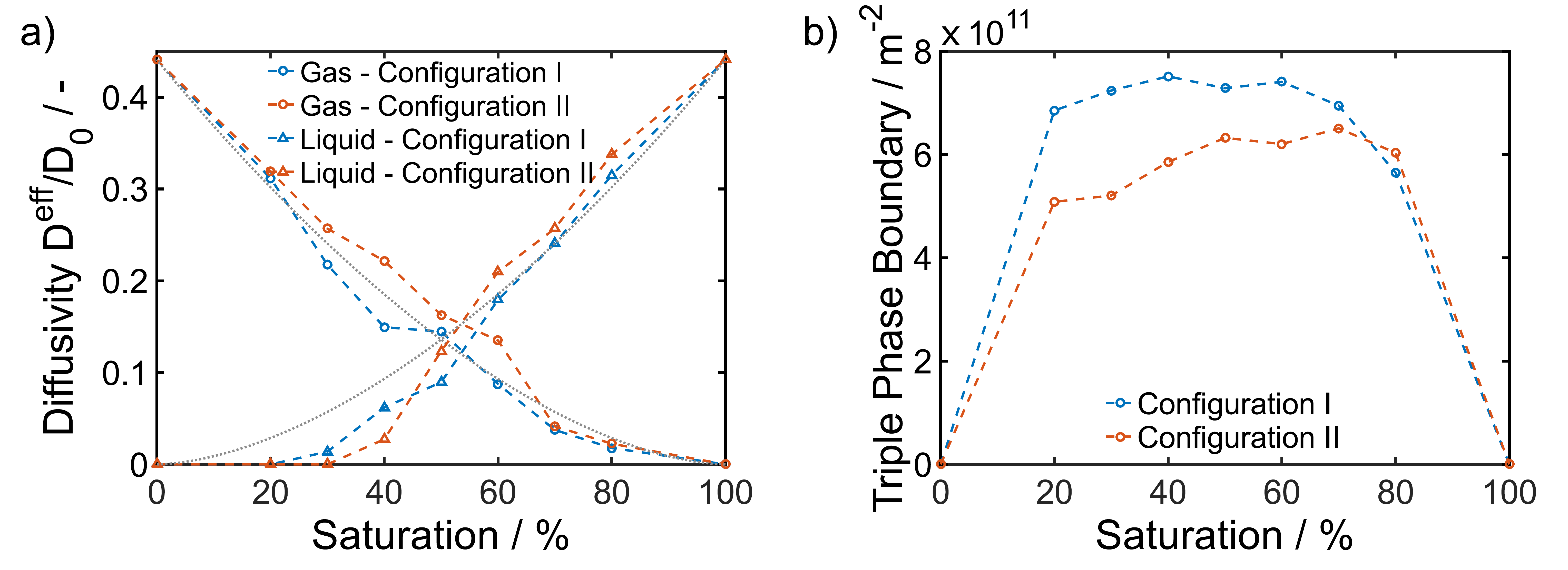}
 \caption{Diffusivity in the main transport direction $y$ (left) and specific length of the triple phase boundary (right) as function of the saturation for configuration I (blue) and configuration II (red). The Bruggeman correlation (Eq. \eqref{eq:methodology_Bruggeman}, $\beta$=1.70) is included as gray line in Figure \ref{fig:results_TransportY} a).}
 \label{fig:results_TransportY}
\end{center}
\end{figure}

Figure \ref{fig:results_Diffusivity} shows the diffusivity in the gas and liquid phase for configuration I (left) and configuration II (right). Results are presented individually for the three spatial coordinates $x$, $y$, and $z$. The graphs show that the transport behavior is anisotropic. Especially, in $z$-direction which is the direction of FIB sampling the transport parameters are smaller. This can be attributed to a lower spatial resolution and indicates that the anisotropy might be an artifact of the reconstruction. The overall trend of the diffusivity with saturation is the same for both initial configurations. The diffusivity of the gas phase decreases with an increasing saturation of the pore space with the liquid phase. This corresponds to the behavior predicted by the Bruggeman correlation (cf. Eq. \eqref{eq:methodology_Bruggeman}) presented in Figure \ref{fig:results_TransportY} a).\\
The main transport direction in the GDE during operation is in our notation denoted by $y$ and we limit our discussion in the remainder of this section to this most important case. Figure \ref{fig:results_TransportY} a) shows the calculated diffusivity in the gas and liquid phase for simulations with initial configuration I and II. As mentioned in Section \ref{sec:results_distribution} the final liquid phase distribution is similar for both cases and, thus, also saturation dependent transport coefficients are comparable. The dotted line represents the Bruggeman correlation with a coefficient of $\beta$=1.7. The graph demonstrates that the correlation gives a reasonable representation of the simulated values although the coefficient is slightly higher than the standard value of 1.5 which is usually assumed in the literature.\\
An important parameter for the performance of the GDE is the length of the triple phase boundary (TPB). Here, gas, liquid, and solid phase are in close contact and reaction kinetics are facile. The dependence of the TPB on the saturation is shown in Figure \ref{fig:results_TransportY} b) for both, configuration I and configuration II. The length of the TPB first increases with saturation. Then, it reaches a plateau because large parts of the pore space are completely filled with electrolyte. The same effect leads to a decrease in TPB length at high saturation. This shows that at a high to moderate saturation the performance of the GDE is improved. The deviation between configuration I and configuration II origins from the more homogeneous distribution of the liquid phase in configuration I.\\
The correlations determined above are important input for simulations on the continuum scale \cite{Horstmann2013:Precipitation, Danner2014:Halfcell} and allow an efficient improvement of gas diffusion electrodes. An overview of this multi-scale approach can also be found in Ref. \cite{Gruebl2014:Methodology}.

\section{Conclusions}
\label{sec:conclusions}
In this article we present a methodology for the characterization of gas diffusion electrodes for metal-air batteries. First, we take FIB-SEM images of an Ag model electrode. In a subsequent step the images are binarized and stacked to a virtual reconstruction of the electrode. On this geometry we performed 2D and 3D LBM simulations with a RK type multiphase model. To the best of our knowledge this study is the first publication investigating the complex multiphase transport mechanism in GDEs using LBM on the pore-scale. An important feature of our model is that we take into account the non-homogeneous wetting behavior of hydrophilic Ag particles and hydrophobic binder. The model is parametrized to represent an aqueous electrolyte system, however, other electrolytes (e.g. organic solutions or ionic liquids) can be treated in the same framework. Results of the simulations are evaluated for the determination of $p_c-s$ curves and saturation dependent transport parameters. Our simulations demonstrate that a series of 2D simulations is an efficient tool for the screening of $p_c-s$ characteristics of GDEs. Transport parameters and specific surface areas are determined from the final phase distributions of 3D simulations. The diffusivity generally follows the trends predicted by the Bruggeman correlation with a coefficient of 1.7. The results of this study provide important input parameters for simulations on the continuum scale. Moreover, the methodology presented here can be used as a tool for the optimization of GDEs for metal air batteries.

\textbf{Acknowledgments}\
Funding from the European Union's innovation programme Horizon 2020 (2014-2020) under grant agreement n\textdegree 646186 (ZAS) is gratefully acknowledged. The authors would also like to thank Norbert Wagner, Institute for Engineering Thermodynamics, DLR Stuttgart for providing the samples of Ag model electrodes and Prasanth Balasubramanian, Ulm University for help in tomogram reconstruction. Santhana Eswara would like to thank J\"org Bernhard and Prof. Ute Kaiser of Ulm University for fruitful discussions. Timo Danner would like to thank Prof. Wolfgang Bessler, University of Applied Sciences Offenburg, for discussions and comments.


\begin{thebibliography}{10}
\expandafter\ifx\csname url\endcsname\relax
  \def\url#1{\texttt{#1}}\fi
\expandafter\ifx\csname urlprefix\endcsname\relax\def\urlprefix{URL }\fi
\expandafter\ifx\csname href\endcsname\relax
  \def\href#1#2{#2} \def\path#1{#1}\fi

\bibitem{Rahman:MetalAirReview}
M.~A. Rahman, X.~Wang, C.~Wen, {High Energy Density Metal-Air Batteries: A
  Review}, J. Electrochem. Soc. 160~(10) (2013) A1759--A1771.
\newblock \href {http://dx.doi.org/10.1149/2.062310jes}
  {\path{doi:10.1149/2.062310jes}}.

\bibitem{Cheng2012:MetalAirReview}
F.~Cheng, J.~Chen, Metal-air batteries: from oxygen reduction electrochemistry
  to cathode catalysts, Chem. Soc. Rev. 41 (2012) 2172--2192.
\newblock \href {http://dx.doi.org/10.1039/C1CS15228A}
  {\path{doi:10.1039/C1CS15228A}}.

\bibitem{Egan:AluminumAirReview}
D.~Egan, C.~{Ponce de Le\'{o}n}, R.~Wood, R.~Jones, K.~Stokes, F.~Walsh,
  {Developments in electrode materials and electrolytes for aluminium-air
  batteries}, J. Power Sources 236 (2013) 293--310.
\newblock \href {http://dx.doi.org/10.1016/j.jpowsour.2013.01.141}
  {\path{doi:10.1016/j.jpowsour.2013.01.141}}.

\bibitem{Girishkumar2010:LiAirPromiseChallenges}
G.~Girishkumar, B.~McCloskey, A.~C. Luntz, S.~Swanson, W.~Wilcke, {Lithium -
  Air Battery: Promise and Challenges}, J. Phys. Chem. Lett. 1~(14) (2010)
  2193--2203.
\newblock \href {http://dx.doi.org/10.1021/jz1005384}
  {\path{doi:10.1021/jz1005384}}.

\bibitem{Zhang2015:MgAir}
T.~Zhang, Z.~Tao, J.~Chen, Magnesium-air batteries: from principle to
  application, Mater. Horiz. 1 (2014) 196--206.
\newblock \href {http://dx.doi.org/10.1039/C3MH00059A}
  {\path{doi:10.1039/C3MH00059A}}.

\bibitem{Palomares:NaReview}
V.~Palomares, P.~Serras, I.~Villaluenga, K.~B. Hueso,
  J.~Carretero-Gonz\'{a}lez, T.~Rojo, {Na-ion batteries, recent advances and
  present challenges to become low cost energy storage systems}, Energy
  Environ. Sci. 5~(3) (2012) 5884.
\newblock \href {http://dx.doi.org/10.1039/c2ee02781j}
  {\path{doi:10.1039/c2ee02781j}}.

\bibitem{Slater:NaReview}
M.~D. Slater, D.~Kim, E.~Lee, C.~S. Johnson, {Sodium-Ion Batteries}, Adv. Func.
  Mater. 23~(8) (2013) 947--958.
\newblock \href {http://dx.doi.org/10.1002/adfm.201200691}
  {\path{doi:10.1002/adfm.201200691}}.

\bibitem{Lee2011:ZnAir}
J.-S. Lee, S.~Tai~Kim, R.~Cao, N.-S. Choi, M.~Liu, K.~T. Lee, J.~Cho, Metal-air
  batteries with high energy density: Li-air versus zn-air, Adv. Energy Mater.
  1~(1) (2011) 34--50.
\newblock \href {http://dx.doi.org/10.1002/aenm.201000010}
  {\path{doi:10.1002/aenm.201000010}}.

\bibitem{Li2014:ZnAir}
Y.~Li, H.~Dai, {Recent advances in zinc-air batteries}, Chem. Soc. Rev. 43~(15)
  (2014) 5257--5275.
\newblock \href {http://dx.doi.org/10.1039/C4CS00015C}
  {\path{doi:10.1039/C4CS00015C}}.

\bibitem{Kim:MetalElectrodesReview}
H.~Kim, G.~Jeong, Y.-U. Kim, J.-H. Kim, C.-M. Park, H.-J. Sohn, {Metallic
  anodes for next generation secondary batteries.}, Chem. Soc. Rev. 42~(23)
  (2013) 9011--34.
\newblock \href {http://dx.doi.org/10.1039/c3cs60177c}
  {\path{doi:10.1039/c3cs60177c}}.

\bibitem{Christensen2012:LiO2Review}
J.~Christensen, P.~Albertus, R.~S. Sanchez-Carrera, T.~Lohmann, B.~Kozinsky,
  R.~Liedtke, J.~Ahmed, A.~Kojic, {A Critical Review of Li/Air Batteries}, J.
  Electrochem. Soc. 159~(2) (2012) R1--R30.
\newblock \href {http://dx.doi.org/10.1149/2.086202jes}
  {\path{doi:10.1149/2.086202jes}}.

\bibitem{ShaoHorn:BridgingMechanisticUnderstanding}
Y.-C. Lu, B.~M. Gallant, D.~G. Kwabi, J.~R. Harding, R.~R. Mitchell, M.~S.
  Whittingham, Y.~Shao-Horn, Lithium-oxygen batteries: bridging mechanistic
  understanding and battery performance, Energy Environ. Sci. 6 (2013)
  750--768.
\newblock \href {http://dx.doi.org/10.1039/C3EE23966G}
  {\path{doi:10.1039/C3EE23966G}}.

\bibitem{Wittmaier2014:Ag}
D.~Wittmaier, N.~Wagner, K.~A. Friedrich, H.~M. Amin, H.~Baltruschat, Modified
  carbon-free silver electrodes for the use as cathodes in lithium-air
  batteries with an aqueous alkaline electrolyte, J. Power Sources 265 (2014)
  299 -- 308.
\newblock \href {http://dx.doi.org/10.1016/j.jpowsour.2014.04.142}
  {\path{doi:10.1016/j.jpowsour.2014.04.142}}.

\bibitem{Wittmaier2014:Ni}
D.~Wittmaier, S.~Aisenbrey, N.~Wagner, K.~A. Friedrich, Bifunctional,
  carbon-free nickel/cobalt-oxide cathodes for lithium-air batteries with an
  aqueous alkaline electrolyte, Electrochim. Acta 149 (2014) 355 -- 363.
\newblock \href {http://dx.doi.org/10.1016/j.electacta.2014.10.088}
  {\path{doi:10.1016/j.electacta.2014.10.088}}.

\bibitem{Wittmaier2015:AgCo}
D.~Wittmaier, N.~A. Ca\~{n}as, I.~Biswas, K.~A. Friedrich, {Highly Stable
  Carbon-Free Ag/Co$_3$O$_4$-Cathodes for Lithium-Air Batteries:
  Electrochemical and Structural Investigations}, Adv. Energy Mater.\href
  {http://dx.doi.org/10.1002/aenm.201500763}
  {\path{doi:10.1002/aenm.201500763}}.

\bibitem{Hirt1981:VOF}
C.~Hirt, B.~Nichols, Volume of fluid (vof) method for the dynamics of free
  boundaries, J. Comput. Phys. 39~(1) (1981) 201 -- 225.
\newblock \href {http://dx.doi.org/10.1016/0021-9991(81)90145-5}
  {\path{doi:10.1016/0021-9991(81)90145-5}}.

\bibitem{Ferreira2015:VOF}
R.~B. Ferreira, D.~S. Falcao, V.~B. Oliveira, A.~M. F.~R. Pinto, {Numerical
  simulations of two-phase flow in proton exchange membrane fuel cells using
  the volume of fluid method - A review}, {J. Power Sources} {277} ({2015})
  {329--342}.
\newblock \href {http://dx.doi.org/10.1016/j.jpowsour.2014.11.124}
  {\path{doi:10.1016/j.jpowsour.2014.11.124}}.

\bibitem{McNamara1988:LBM}
G.~R. McNamara, G.~Zanetti, Use of the boltzmann equation to simulate
  lattice-gas automata, Phys. Rev. Lett. 61 (1988) 2332--2335.
\newblock \href {http://dx.doi.org/10.1103/PhysRevLett.61.2332}
  {\path{doi:10.1103/PhysRevLett.61.2332}}.

\bibitem{Sukop2006:LBM}
M.~C. Sukop, D.~T. Thorne, Lattice Boltzmann Modeling: An Introduction for
  Geoscientists and Engineers, 1st Edition, Springer Berlin Heidelberg, 2006.

\bibitem{Succi2001:LBM}
S.~Succi, The Lattice Boltzmann Equation: For Fluid Dynamics and Beyond,
  Numerical Mathematics and Scientific Computation, Clarendon Press, 2001.

\bibitem{Sukop2015:MultiphaseLBM}
H.~Huang, M.~Sukop, X.~Lu, Multiphase Lattice Boltzmann Methods: Theory and
  Application, Wiley, 2015.

\bibitem{Shan1993:SC1}
X.~Shan, H.~Chen, Lattice boltzmann model for simulating flows with multiple
  phases and components, Phys. Rev. E 47 (1993) 1815--1819.
\newblock \href {http://dx.doi.org/10.1103/PhysRevE.47.1815}
  {\path{doi:10.1103/PhysRevE.47.1815}}.

\bibitem{Shan1993:SC2}
X.~Shan, H.~Chen, Simulation of nonideal gases and liquid-gas phase transitions
  by the lattice boltzmann equation, Phys. Rev. E 49 (1994) 2941--2948.
\newblock \href {http://dx.doi.org/10.1103/PhysRevE.49.2941}
  {\path{doi:10.1103/PhysRevE.49.2941}}.

\bibitem{Swift1995:FE1}
M.~R. Swift, W.~R. Osborn, J.~M. Yeomans, Lattice boltzmann simulation of
  nonideal fluids, Phys. Rev. Lett. 75 (1995) 830--833.
\newblock \href {http://dx.doi.org/10.1103/PhysRevLett.75.830}
  {\path{doi:10.1103/PhysRevLett.75.830}}.

\bibitem{Swift1996:FE2}
M.~R. Swift, E.~Orlandini, W.~R. Osborn, J.~M. Yeomans, Lattice boltzmann
  simulations of liquid-gas and binary fluid systems, Phys. Rev. E 54 (1996)
  5041--5052.
\newblock \href {http://dx.doi.org/10.1103/PhysRevE.54.5041}
  {\path{doi:10.1103/PhysRevE.54.5041}}.

\bibitem{Rothman1988:RK}
D.~H. Rothman, J.~M. Keller, Immiscible cellular-automaton fluids, J. Stat.
  Phys. 52~(3-4) (1988) 1119--1127.
\newblock \href {http://dx.doi.org/10.1007/BF01019743}
  {\path{doi:10.1007/BF01019743}}.

\bibitem{Gunstensen1991:RK1}
A.~K. Gunstensen, D.~H. Rothman, S.~Zaleski, G.~Zanetti, Lattice boltzmann
  model of immiscible fluids, Phys. Rev. A 43 (1991) 4320--4327.
\newblock \href {http://dx.doi.org/10.1103/PhysRevA.43.4320}
  {\path{doi:10.1103/PhysRevA.43.4320}}.

\bibitem{Gunstensen1992:RK2}
A.~K. Gunstensen, D.~H. Rothman, Microscopic modeling of immiscible fluids in
  three dimensions by a lattice boltzmann method, Europhys. Lett. 18~(2) (1992)
  157.

\bibitem{Inamuro2004:FE1}
T.~Inamuro, T.~Ogata, S.~Tajima, N.~Konishi, {A lattice Boltzmann method for
  incompressible two-phase flows with large density differences}, J. Comput.
  Phys. 198~(2) (2004) 628--644.
\newblock \href {http://dx.doi.org/10.1016/j.jcp.2004.01.019}
  {\path{doi:10.1016/j.jcp.2004.01.019}}.

\bibitem{Zheng2006}
H.~Zheng, C.~Shu, Y.~Chew, {A lattice Boltzmann model for multiphase flows with
  large density ratio}, J. Comput. Phys. 218~(1) (2006) 353--371.
\newblock \href {http://dx.doi.org/10.1016/j.jcp.2006.02.015}
  {\path{doi:10.1016/j.jcp.2006.02.015}}.

\bibitem{Niu2007:LBM}
X.~Niu, T.~Munekata, S.~Hyodo, K.~Suga, An investigation of water-gas transport
  processes in the gas-diffusion-layer of a PEM fuel cell by a multiphase
  multiple-relaxation-time lattice boltzmann model, J. Power Sources 172~(2)
  (2007) 542 -- 552.
\newblock \href {http://dx.doi.org/10.1016/j.jpowsour.2007.05.081}
  {\path{doi:10.1016/j.jpowsour.2007.05.081}}.

\bibitem{Park2008:LBM}
J.~Park, X.~Li, {Multi-phase micro-scale flow simulation in the electrodes of a
  PEM fuel cell by lattice Boltzmann method}, J. Power Sources 178~(1) (2008)
  248--257.
\newblock \href {http://dx.doi.org/10.1016/j.jpowsour.2007.12.008}
  {\path{doi:10.1016/j.jpowsour.2007.12.008}}.

\bibitem{Mukherjee2011:Review}
P.~P. Mukherjee, Q.~Kang, C.-Y. Wang, {Pore-scale modeling of two-phase
  transport in polymer electrolyte fuel cells-progress and perspective},
  {Energy Environ. Sci.} {4}~({2}) ({2011}) {346--369}.
\newblock \href {http://dx.doi.org/10.1039/b926077c}
  {\path{doi:10.1039/b926077c}}.

\bibitem{Hao2012:CapillaryPressure}
L.~Hao, P.~Cheng, {Capillary pressures in carbon paper gas diffusion layers
  having hydrophilic and hydrophobic pores}, Int. J. Heat Mass Trans. 55~(1-3)
  (2012) 133--139.
\newblock \href {http://dx.doi.org/10.1016/j.ijheatmasstransfer.2011.08.049}
  {\path{doi:10.1016/j.ijheatmasstransfer.2011.08.049}}.

\bibitem{Nabovati2014:PorosityEffect}
A.~Nabovati, J.~Hinebaugh, A.~Bazylak, C.~H. Amon, {Effect of porosity
  heterogeneity on the permeability and tortuosity of gas diffusion layers in
  polymer electrolyte membrane fuel cells}, {J. Power Sources} {248} ({2014})
  {83--90}.
\newblock \href {http://dx.doi.org/10.1016/j.jpowsour.2013.09.061}
  {\path{doi:10.1016/j.jpowsour.2013.09.061}}.

\bibitem{Dong2015:LBM_compression}
D.~H. Jeon, H.~Kim, {Effect of compression on water transport in gas diffusion
  layer of polymer electrolyte membrane fuel cell using lattice Boltzmann
  method}, {J. Power Sources} {294} ({2015}) {393--405}.
\newblock \href {http://dx.doi.org/10.1016/j.jpowsour.2015.06.080}
  {\path{doi:10.1016/j.jpowsour.2015.06.080}}.

\bibitem{Leclaire2011:ColorGradient}
S.~Leclaire, M.~Reggio, J.-Y. Tr\'{e}panier, {Isotropic color gradient for
  simulating very high-density ratios with a two-phase flow lattice Boltzmann
  model}, Comput. Fluids 48~(1) (2011) 98--112.
\newblock \href {http://dx.doi.org/10.1016/j.compfluid.2011.04.001}
  {\path{doi:10.1016/j.compfluid.2011.04.001}}.

\bibitem{Leclaire2012:Recoloring}
S.~Leclaire, M.~Reggio, J.-Y. Tr\'{e}panier, {Numerical evaluation of two
  recoloring operators for an immiscible two-phase flow lattice Boltzmann
  model}, Appl. Math. Model. 36~(5) (2012) 2237 -- 2252.
\newblock \href {http://dx.doi.org/10.1016/j.apm.2011.08.027}
  {\path{doi:10.1016/j.apm.2011.08.027}}.

\bibitem{Leclaire2013:MultiPhase}
S.~Leclaire, M.~Reggio, J.-Y. Tr\'{e}panier, {Progress and investigation on
  lattice Boltzmann modeling of multiple immiscible fluids or components with
  variable density and viscosity ratios}, J. Comput. Phys. 246 (2013) 318--342.
\newblock \href {http://dx.doi.org/10.1016/j.jcp.2013.03.039}
  {\path{doi:10.1016/j.jcp.2013.03.039}}.

\bibitem{Liu2012:RK3D}
H.~Liu, A.~J. Valocchi, Q.~Kang, {Three-dimensional lattice Boltzmann model for
  immiscible two-phase flow simulations}, Phys. Rev. E 85~(4) (2012) 046309.
\newblock \href {http://dx.doi.org/10.1103/PhysRevE.85.046309}
  {\path{doi:10.1103/PhysRevE.85.046309}}.

\bibitem{Molaeimanesh:PTFE}
G.~R. Molaeimanesh, M.~H. Akbari, {Impact of PTFE distribution on the removal
  of liquid water from a PEMFC electrode by lattice Boltzmann method}, {Int. J.
  Hydrogen Energy} {39}~({16}) ({2014}) {8401--8409}.
\newblock \href {http://dx.doi.org/10.1016/j.ijhydene.2014.03.089}
  {\path{doi:10.1016/j.ijhydene.2014.03.089}}.

\bibitem{Hao2010:Permeabilities}
L.~Hao, P.~Cheng, {Pore-scale simulations on relative permeabilities of porous
  media by lattice Boltzmann method}, {Int. J. Heat Mass Tran.} {53}~({9-10})
  ({2010}) {1908--1913}.
\newblock \href {http://dx.doi.org/10.1016/j.ijheatmasstransfer.2009.12.066}
  {\path{doi:10.1016/j.ijheatmasstransfer.2009.12.066}}.

\bibitem{Hao2010:WaterTransport}
L.~Hao, P.~Cheng, {Lattice Boltzmann simulations of water transport in gas
  diffusion layer of a polymer electrolyte membrane fuel cell}, {J. Power
  Sources} {195}~({12, SI}) ({2010}) {3870--3881}.
\newblock \href {http://dx.doi.org/10.1016/j.jpowsour.2009.11.125}
  {\path{doi:10.1016/j.jpowsour.2009.11.125}}.

\bibitem{Zhou2010:WaterTransport}
P.~Zhou, C.~W. Wu, {Liquid water transport mechanism in the gas diffusion
  layer}, {J. Power Sources} {195}~({5}) ({2010}) {1408--1415}.
\newblock \href {http://dx.doi.org/10.1016/j.jpowsour.2009.09.019}
  {\path{doi:10.1016/j.jpowsour.2009.09.019}}.

\bibitem{Mukherjee2009:WaterManagement}
P.~P. Mukherjee, C.-Y. Wang, Q.~Kang, {Mesoscopic modeling of two-phase
  behavior and flooding phenomena in polymer electrolyte fuel cells},
  {Electrochim. Acta} {54}~({27}) ({2009}) {6861--6875}.
\newblock \href {http://dx.doi.org/10.1016/j.electacta.2009.06.066}
  {\path{doi:10.1016/j.electacta.2009.06.066}}.

\bibitem{Danner2014:Halfcell}
T.~Danner, B.~Horstmann, D.~Wittmaier, N.~Wagner, W.~G. Bessler, {Reaction and
  transport in Ag/Ag$_2$O gas diffusion electrodes of aqueous Li-O$_2$
  batteries: Experiments and modeling}, J. Power Sources 264~(0) (2014) 320 --
  332.
\newblock \href {http://dx.doi.org/10.1016/j.jpowsour.2014.03.149}
  {\path{doi:10.1016/j.jpowsour.2014.03.149}}.

\bibitem{Eswara2014:FIB-SEM}
S.~K. Eswara-Moorthy, P.~Balasubramanian, W.~van Mierlo, J.~Bernhard,
  M.~Marinaro, M.~Wohlfahrt-Mehrens, L.~J\"orissen, U.~Kaiser, An in situ
  sem-fib-based method for contrast enhancement and tomographic reconstruction
  for structural quantification of porous carbon electrodes, Microsc.
  Microanal. 20 (2014) 1576--1580.
\newblock \href {http://dx.doi.org/10.1017/S1431927614012884}
  {\path{doi:10.1017/S1431927614012884}}.

\bibitem{GeoDict:UserGuide}
{Math2Market GmbH - Becker, J.; Wiegmann, A.},
  \href{http://www.geodict.com}{{GeoDict}} (2013).
\newline\urlprefix\url{http://www.geodict.com}

\bibitem{Tolke2002:RK}
J.~T\"{o}lke, M.~Krafczyk, M.~Schulz, E.~Rank, {Lattice Boltzmann simulations
  of binary fluid flow through porous media.}, Philos. T. Roy. Soc. A
  360~(1792) (2002) 535--45.
\newblock \href {http://dx.doi.org/10.1098/rsta.2001.0944}
  {\path{doi:10.1098/rsta.2001.0944}}.

\bibitem{Reis2007:RK}
T.~Reis, T.~N. Phillips, {Lattice Boltzmann model for simulating immiscible
  two-phase flows}, J. Phys. A-Math. Theor. 40~(14) (2007) 4033--4053.
\newblock \href {http://dx.doi.org/10.1088/1751-8113/40/14/018}
  {\path{doi:10.1088/1751-8113/40/14/018}}.

\bibitem{Bhatnagar1954:BGKOperator}
P.~L. Bhatnagar, E.~P. Gross, M.~Krook, A model for collision processes in
  gases. i. small amplitude processes in charged and neutral one-component
  systems, Phys. Rev. 94 (1954) 511--525.
\newblock \href {http://dx.doi.org/10.1103/PhysRev.94.511}
  {\path{doi:10.1103/PhysRev.94.511}}.

\bibitem{Sbragaglia2007}
M.~Sbragaglia, R.~Benzi, L.~Biferale, S.~Succi, K.~Sugiyama, F.~Toschi,
  {Generalized lattice Boltzmann method with multirange pseudopotential}, Phys.
  Rev. E 75~(2) (2007) 026702.
\newblock \href {http://dx.doi.org/10.1103/PhysRevE.75.026702}
  {\path{doi:10.1103/PhysRevE.75.026702}}.

\bibitem{Latva-Kokko2005:LatticePinning}
M.~Latva-Kokko, D.~Rothman, {Diffusion properties of gradient-based lattice
  Boltzmann models of immiscible fluids}, Phys. Rev. E 71~(5) (2005) 056702.
\newblock \href {http://dx.doi.org/10.1103/PhysRevE.71.056702}
  {\path{doi:10.1103/PhysRevE.71.056702}}.

\bibitem{Latva-Kokko2005:ContactAngle}
M.~Latva-Kokko, D.~Rothman, {Static contact angle in lattice Boltzmann models
  of immiscible fluids}, Phys. Rev. E 72~(4) (2005) 046701.
\newblock \href {http://dx.doi.org/10.1103/PhysRevE.72.046701}
  {\path{doi:10.1103/PhysRevE.72.046701}}.

\bibitem{Dwenger2011:CapillaryPressure}
S.~Dwenger, G.~Eigenberger, U.~Nieken, {Measurement of Capillary
  Pressure--Saturation Relationships Under Defined Compression Levels for Gas
  Diffusion Media of PEM Fuel Cells}, Transport Porous Med. 91~(1) (2011)
  281--294.
\newblock \href {http://dx.doi.org/10.1007/s11242-011-9844-4}
  {\path{doi:10.1007/s11242-011-9844-4}}.

\bibitem{Schaap2007:Rescaling}
M.~G. Schaap, M.~L. Porter, B.~S.~B. Christensen, D.~Wildenschild, {Comparison
  of pressure-saturation characteristics derived from computed tomography and
  lattice Boltzmann simulations}, Water Resour. Res. 43~(12) (2007) W12S06.
\newblock \href {http://dx.doi.org/10.1029/2006WR005730}
  {\path{doi:10.1029/2006WR005730}}.

\bibitem{Zhao:ContactAngleAgPTFE}
Q.~Zhao, Y.~Liu, C.~Wang, {Development and evaluation of electroless Ag-PTFE
  composite coatings with anti-microbial and anti-corrosion properties}, Appl.
  Surf. Sci. 252~(5) (2005) 1620--1627.
\newblock \href {http://dx.doi.org/10.1016/j.apsusc.2005.02.098}
  {\path{doi:10.1016/j.apsusc.2005.02.098}}.

\bibitem{Gostick2006:CapillaryPressure}
J.~T. Gostick, M.~W. Fowler, M.~A. Ioannidis, M.~D. Pritzker, Y.~Volfkovich,
  A.~Sakars, {Capillary pressure and hydrophilic porosity in gas diffusion
  layers for polymer electrolyte fuel cells}, J. Power Sources 156~(2) (2006)
  375--387.
\newblock \href {http://dx.doi.org/10.1016/j.jpowsour.2005.05.086}
  {\path{doi:10.1016/j.jpowsour.2005.05.086}}.

\bibitem{Kumbur2007:LeverettIII}
E.~C. Kumbur, K.~V. Sharp, M.~M. Mench, {Validated Leverett Approach for
  Multiphase Flow in PEFC Diffusion Media - III. Temperature Effect and Unified
  Approach}, J. Electrochem. Soc. 154~(12) (2007) B1315 -- B1324.
\newblock \href {http://dx.doi.org/10.1149/1.2784286}
  {\path{doi:10.1149/1.2784286}}.

\bibitem{Fairweather2007:CapillaryPressure}
J.~D. Fairweather, P.~Cheung, J.~St-Pierre, D.~T. Schwartz, {A microfluidic
  approach for measuring capillary pressure in PEMFC gas diffusion layers},
  Electrochem. Commun. 9~(9) (2007) 2340--2345.
\newblock \href {http://dx.doi.org/10.1016/j.elecom.2007.06.042}
  {\path{doi:10.1016/j.elecom.2007.06.042}}.

\bibitem{Gostick2008:CapillaryPressure}
J.~T. Gostick, M.~A. Ioannidis, M.~W. Fowler, M.~D. Pritzker, Direct
  measurement of the capillary pressure characteristics of water-air-gas
  diffusion layer systems for pem fuel cells, Electrochem. Commun. 10~(10)
  (2008) 1520 -- 1523.
\newblock \href {http://dx.doi.org/10.1016/j.elecom.2008.08.008}
  {\path{doi:10.1016/j.elecom.2008.08.008}}.

\bibitem{Nguyen2008:CapillaryPressure}
T.~V. Nguyen, G.~Lin, H.~Ohn, X.~Wang, {Measurement of Capillary Pressure
  Property of Gas Diffusion Media Used in Proton Exchange Membrane Fuel Cells},
  Electrochem. Solid St. 11~(8) (2008) B127--B131.
\newblock \href {http://dx.doi.org/10.1149/1.2929063}
  {\path{doi:10.1149/1.2929063}}.

\bibitem{Leverett1940:J}
M.~Leverett, Capillary behavior in porous solids, T. Am. I. Min. Met. Eng. 142
  (1941) 152--169.
\newblock \href {http://dx.doi.org/10.2118/941152-G}
  {\path{doi:10.2118/941152-G}}.

\bibitem{Udell1985:J}
K.~S. Udell, Heat transfer in porous media considering phase change and
  capillarity-the heat pipe effect, Int. J. Heat Mass Tran. 28~(2) (1985) 485
  -- 495.
\newblock \href {http://dx.doi.org/10.1016/0017-9310(85)90082-1}
  {\path{doi:10.1016/0017-9310(85)90082-1}}.

\bibitem{Garcia2015:Diffusivity}
P.~A. Garcia-Salaberri, J.~T. Gostick, G.~Hwang, A.~Z. Weber, M.~Vera,
  Effective diffusivity in partially-saturated carbon-fiber gas diffusion
  layers: Effect of local saturation and application to macroscopic continuum
  models, J. Power Sources 296 (2015) 440 -- 453.
\newblock \href {http://dx.doi.org/10.1016/j.jpowsour.2015.07.034}
  {\path{doi:10.1016/j.jpowsour.2015.07.034}}.

\bibitem{Horstmann2013:Precipitation}
B.~Horstmann, T.~Danner, W.~G. Bessler, {Precipitation in aqueous
  lithium--oxygen batteries: a model-based analysis}, Energy Environ. Sci. 6
  (2013) 1299--1314.
\newblock \href {http://dx.doi.org/10.1039/c3ee24299d}
  {\path{doi:10.1039/c3ee24299d}}.

\bibitem{Gruebl2014:Methodology}
D.~Gr\"{u}bl, T.~Danner, V.~P. Schulz, A.~Latz, W.~G. Bessler,
  Multi-methodology modeling and design of lithium-air cells with aqueous
  electrolyte, ECS Trans. 62~(1) (2014) 137--149.
\newblock \href {http://dx.doi.org/10.1149/06201.0137ecst}
  {\path{doi:10.1149/06201.0137ecst}}.

\bibitem{Qian1992:Nomencalture}
Y.~Qian, D.~D'Humi\`{e}res, P.~Lallemand, {Lattice BGK Models for Navier-Stokes
  Equation }, {Europhys. Lett.} {17}~({6BIS}) ({1992}) {479--484}.
\newblock \href {http://dx.doi.org/10.1209/0295-5075/17/6/001}
  {\path{doi:10.1209/0295-5075/17/6/001}}.

\end{thebibliography}

\clearpage
\listoftables
\clearpage

\begin{table}[h!]
\begin{center}
\begin{tabular}{l|c|c|c}
 & Dimensions / \textmu m & Voxels / - & Voxel size / nm \\
 & x/y/z & x/y/z & x/y/z \\ \hline
FIB-SEM & 27.8/23.9/10.0 & 504/346/84 & 55.25/69.1l/119 \\ \hline
Reconstruction & 27.8/23.9/10.0 & 504/432/180 & 55.25 \\ \hline
LBM & 27.8/23.9/20.0 & 126/108/90 & 221 \\ \hline
\hline
\end{tabular}
\caption{Dimensions and resolution of the electrode sample in $x, y, z$ direction. The reconstructed electrodes are based on cubic voxels. For the LBM simulations the structure is mirrored in $z$ direction to increase the computational domain.}
\label{tab:recon_StrucData}
\end{center}
\end{table}

\begin{table}[h!]
\resizebox{\columnwidth}{!}{%
\begin{tabular}{l|c|c|c|c|c|c|c}
 & \multicolumn{ 2}{c|}{$\rho$} & \multicolumn{ 2}{c|}{$\nu$} & $\sigma$ & \multicolumn{ 2}{c}{$\Theta$} \\ \cline{ 2- 8}
 & Gas & Liquid & Gas & Liquid & Gas-Liquid & Ag & Binder \\ \hline
Physical & 1.18 & 997 & 1.58$\cdot 10^{-5}$ & 8.93$\cdot 10^{-7}$ & 7.28$\cdot 10^{-2}$ & 67 \textdegree \cite{Zhao:ContactAngleAgPTFE} & 140 \textdegree \cite{Mukherjee2009:WaterManagement} \\ \hline
LBM & 1 & 1 &  1/6 &  1/6 & 0.1 & 1.195/0.805\textdagger & 0.357/1.643\textdagger \\ \hline
\hline
\multicolumn{8}{l}{\textdagger $\;$ Density on wall nodes (gas/liquid) mu lu$^{-3}$}
\end{tabular}
}
\caption{Physical parameters of the system (SI units) and corresponding input parameters of the LBM simulations (lattice units).}
\label{tab:LBM_Parameters}
\end{table}

\begin{table}[h!]
\begin{center}
\begin{tabular}{l|c|c|c|c|c}
            & $A$    & $B$                   & $C$   & $D$   & $E$ \\ \hline
 configuration I   & -0.813 & 0.233 & 5.079 & 2.323$\cdot$10$^{-3}$ & 13.47 \\ \hline
 configuration II & 4.557$\cdot$10$^{-2}$  & 3.797$\cdot$10$^{-2}$ & 8.138 & 0.01 & 10.98 \\ \hline
\hline
\end{tabular}
\caption{Parameters of the Leverett function resulting from a fit to Eq. \eqref{eq:results_LeverettFit}.}
\label{tab:results_LeverettParameters}
 \end{center}
\end{table}

\clearpage
\listoffigures
\clearpage

\captionsetup[figure]{list=no}
\setcounter{figure}{0}

\begin{figure}[ht!]
  \includegraphics[width=\textwidth,keepaspectratio=true]{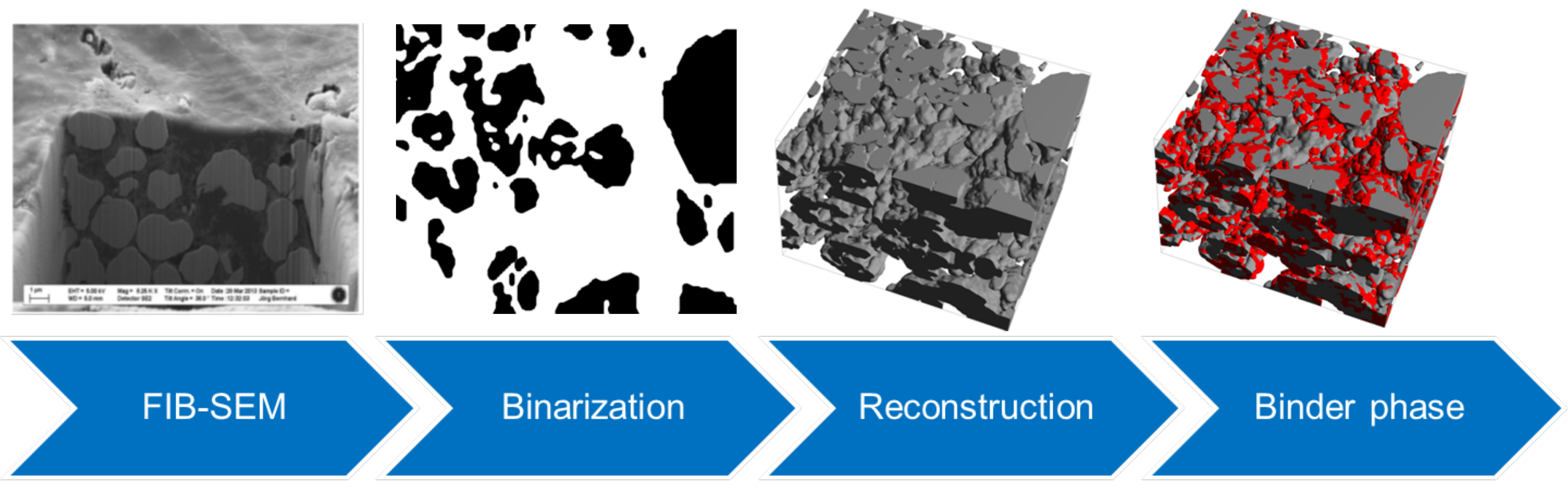}
 \caption{Methodology to reconstruct the LBM simulation domain from FIB-SEM images. Full details regarding the dimensions and voxel sizes are given in Table \ref{tab:recon_StrucData}. Scale bar of SEM image is 1 \textmu m.}
\end{figure}%

\begin{figure}[hb!]
 \includegraphics[width=\textwidth,keepaspectratio=true]{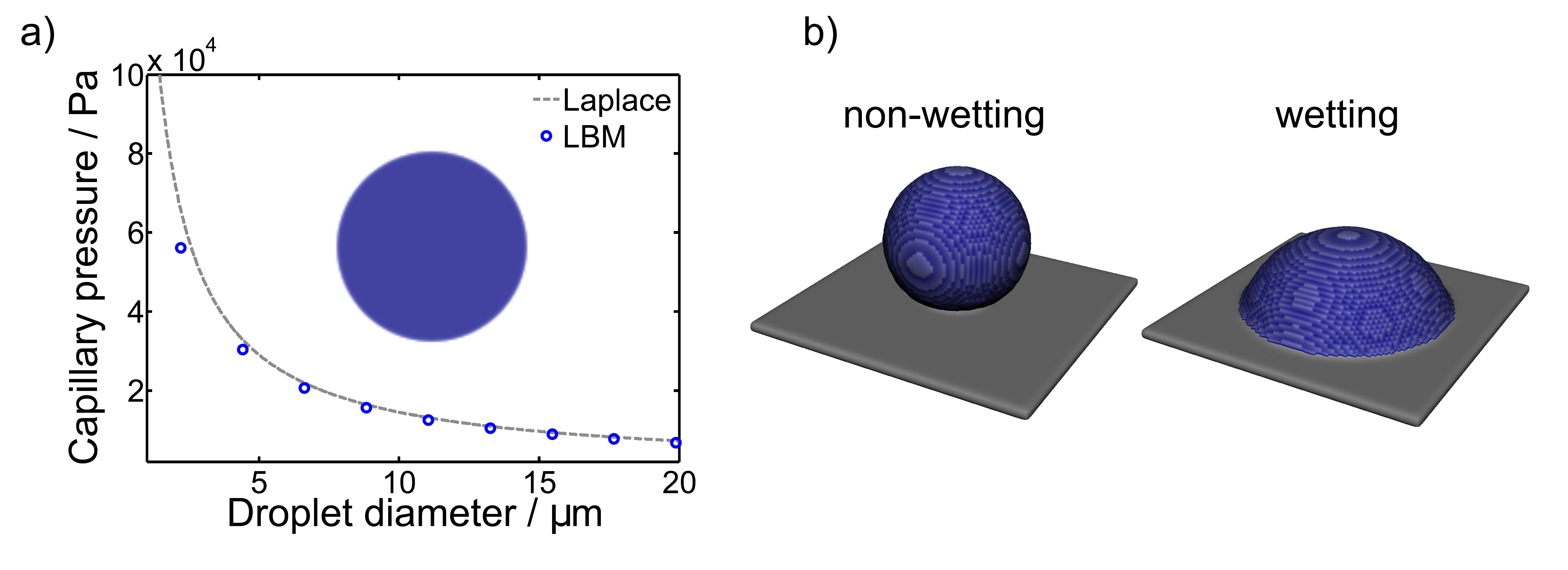}
 \caption{Validation simulations of the LBM model. Left: Bubble test. Right: Influence of non-wetting and wetting surfaces.}
\end{figure}%

\begin{figure}[ht!]
  \includegraphics[width=0.9\textwidth,keepaspectratio=true]{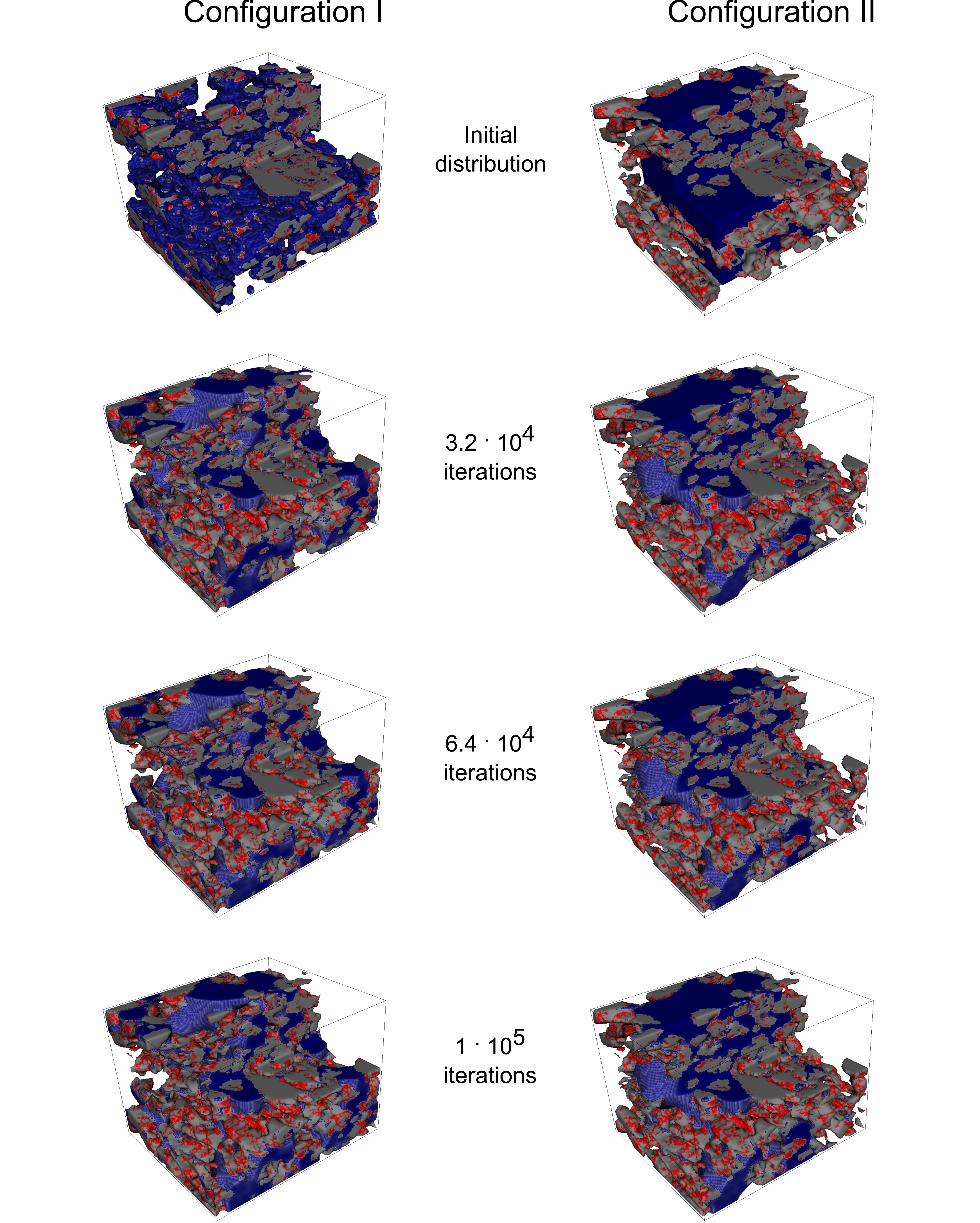}
 \caption{Electrolyte distribution in the porous GDE for configuration I (left) and configuration II (right).}
\end{figure}%

\begin{figure}[ht!]
 \includegraphics[width=\textwidth,keepaspectratio=true]{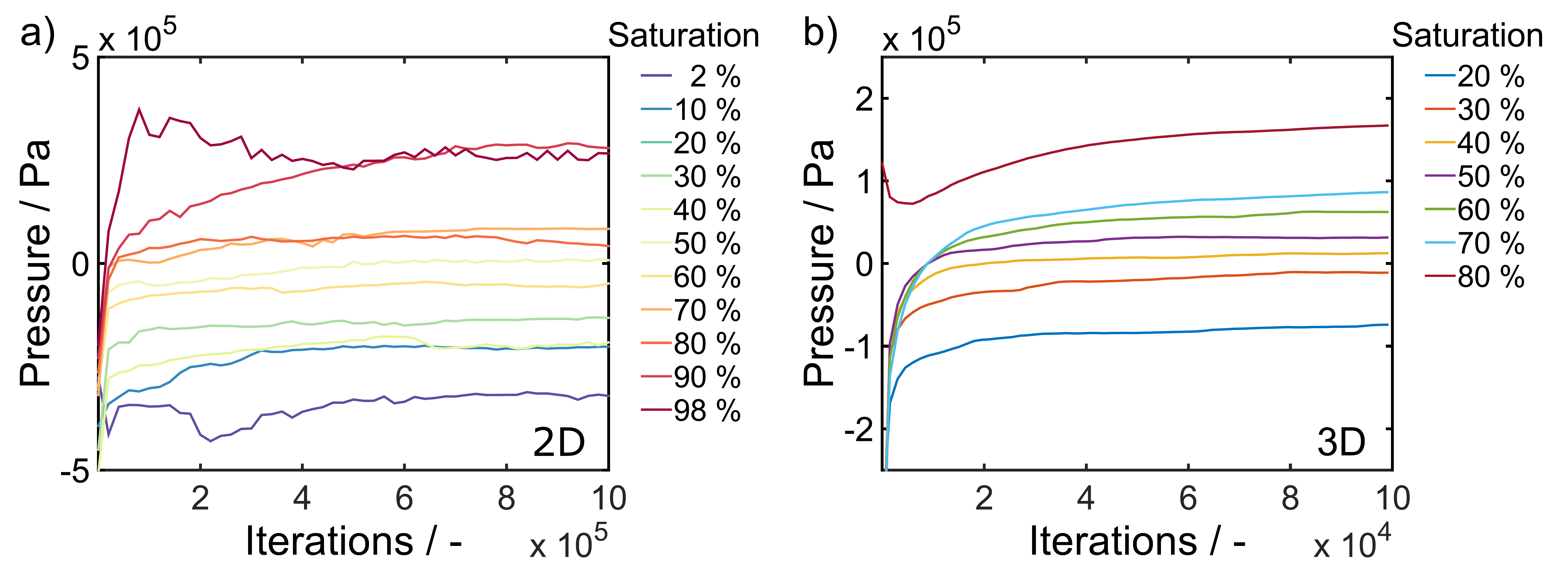}
 \caption{Pressure as function of iterations during simulations employing configuration I. Left: 2D simulations. Right: 3D simulations.}
\end{figure}%

\begin{figure}[hb!]
 \includegraphics[width=\textwidth,keepaspectratio=true]{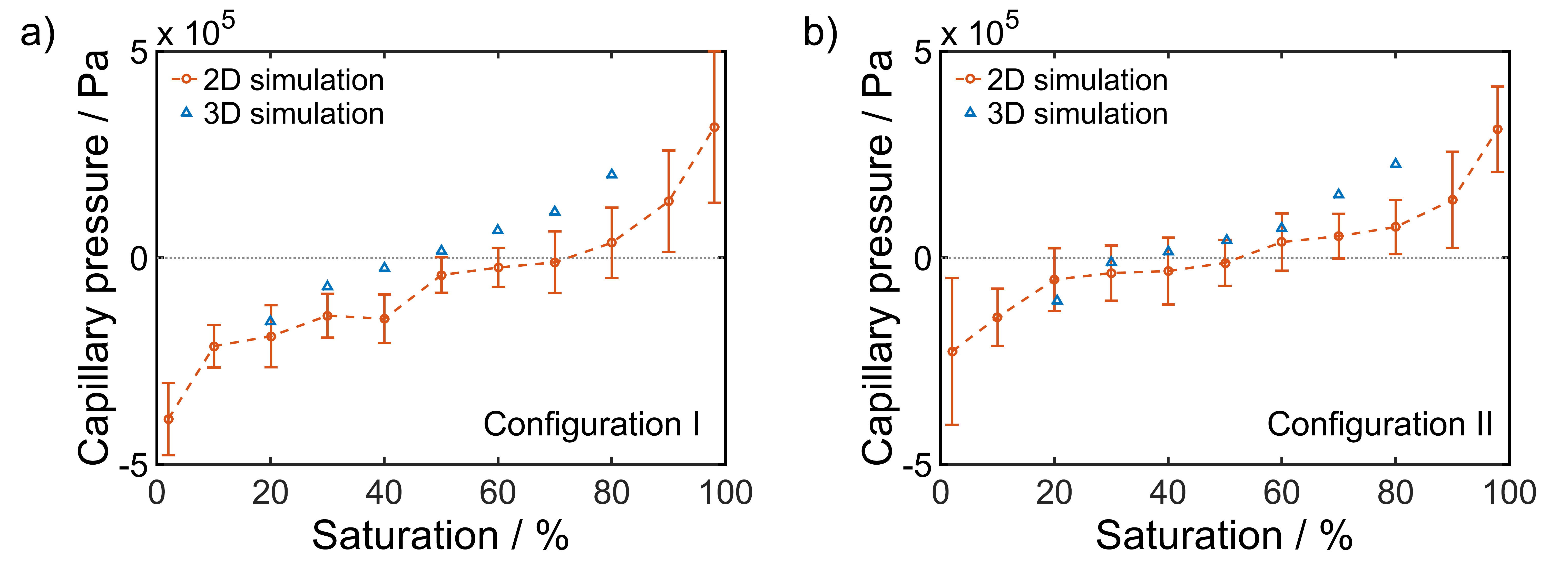}
 \caption{Pressure-saturation curves during for configuration I (left) and configuration II (right). Results of the 2D simulations are shown in red color and error bars represent the standard deviation of the simulations. Results of the 3D simulations are represented by blue triangles.}
\end{figure}%

\begin{figure}[ht!]
\begin{center}
 \includegraphics[width=0.5\textwidth,keepaspectratio=true]{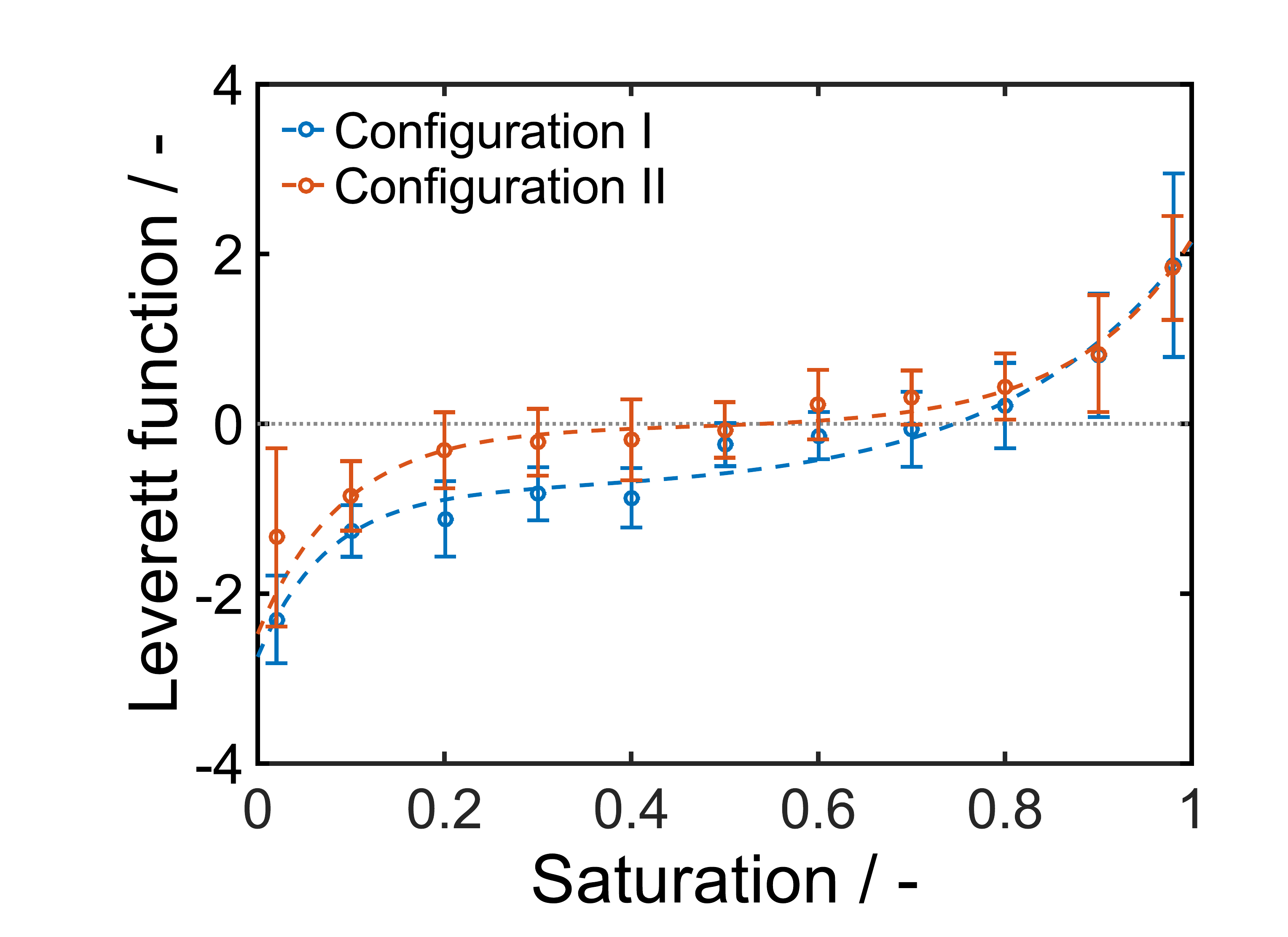}
 \caption{Plot of the dimensionless Leverett Function over saturation for configuration I (blue) and configuration II (red). Dashed lines are a result of the fit to Eq. \eqref{eq:results_LeverettFit}. Symbols represent corresponding simulation results.}
 \end{center}
\end{figure}%

\begin{figure}[hb!]
 \includegraphics[width=\textwidth,keepaspectratio=true]{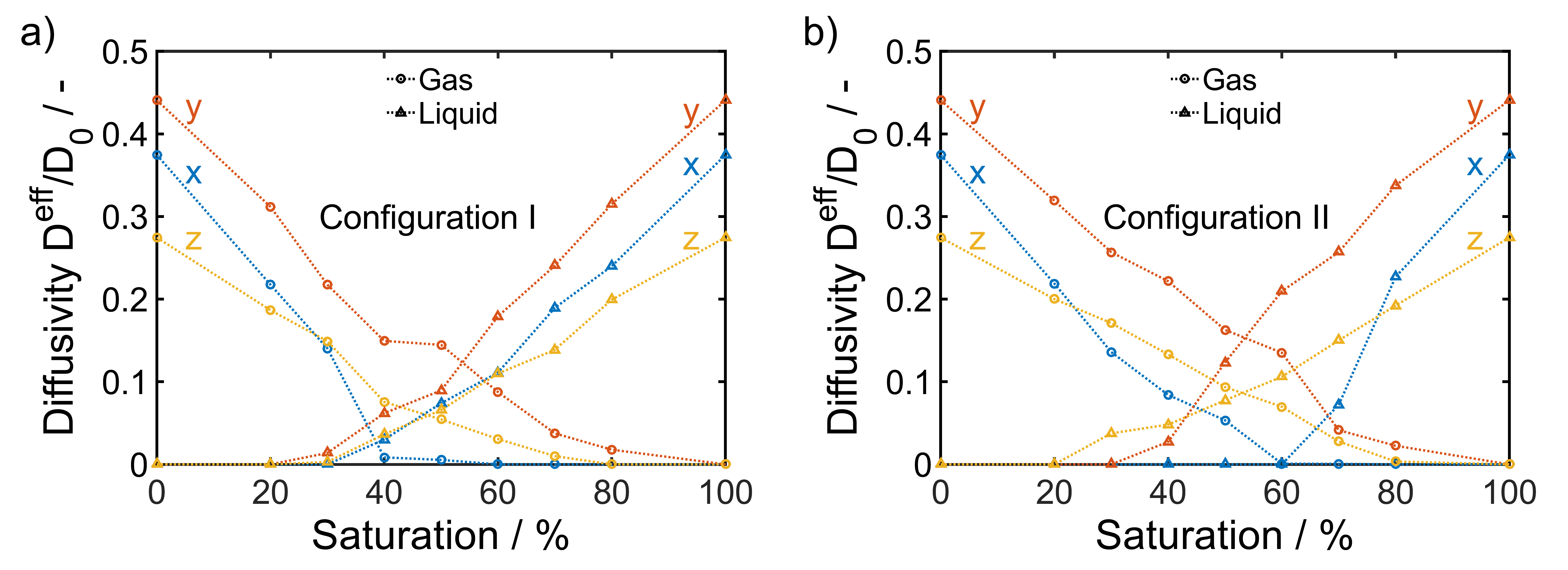}
 \caption{Diffusivity in the gas (open circles) and liquid (open triangles) phase. Values are calculated based on the final electrolyte distribution obtained in the 3D LBM simulations. Plots for the three spatial coordinates ($x$,$y$,$z$) are shown separately for configuration I (left) and configuration II (right).}
\end{figure}%

\begin{figure}[ht!]
\begin{center}
 \includegraphics[width=\textwidth,keepaspectratio=true]{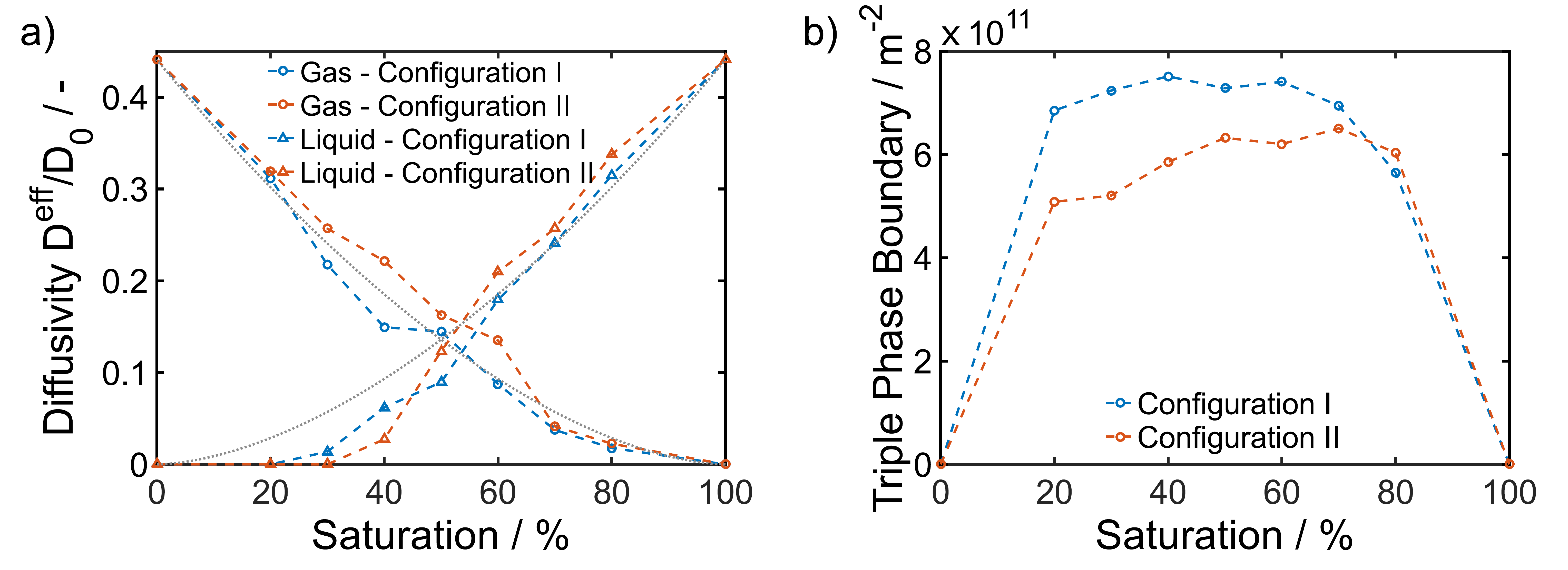}
 \caption{Diffusivity in the main transport direction $y$ (left) and specific length of the triple phase boundary (right) as function of the saturation for configuration I (blue) and configuration II (red). The Bruggeman correlation (Eq. \eqref{eq:methodology_Bruggeman}, $\beta$=1.70) is included as gray line in Figure \ref{fig:results_TransportY} a).}
\end{center}
\end{figure}%

\clearpage
\appendix
\section{LBM parameters} 
\label{sec:appendix_parameters}

\subsection{Computational lattice}

The D2Q9 and D3Q19 lattices \cite{Qian1992:Nomencalture} in cartesian coordinates are given by 
\begin{equation} 
\mathbf{e} = c
\scalemath{0.75}{
  \begin{pmatrix}
    0 & 1 & 0 & -1 &  0 & 1 & -1 &  1 & -1 \\ 
    0 & 0 & 1 &  0 & -1 & 1 &  1 & -1 & -1 
  \end{pmatrix}
  }
  \label{eq:appendix_D2Q9Lattice}
\end{equation}
and 
\begin{equation} 
\mathbf{e} = c
\scalemath{0.75}{
  \begin{pmatrix}
    0 & 1 & -1 & 0 & 0  & 0 & 0  & 1 & -1 & 1  & -1 & 1 & -1 & 1 & -1 & 0 & 0 & 0 & 0 \\
    0 & 0 & 0  & 1 & -1 & 0 & 0  & 1 & -1 & -1 & 1  & 0 & 0  & 0  & 0  & 1 & -1 & 1  & -1\\
    0 & 0 & 0  & 0 & 0  & 1 & -1 & 0 & 0  & 0  & 0  & 1 & -1 & -1 & 1  & 1 & -1 & -1 & 1
  \end{pmatrix}
  }
  \;,
  \label{eq:appendix_D3Q19Lattice}
\end{equation}
respectively.

\subsection{Single-phase collision}
For the D2Q9 and D3Q19 lattices $w_i$ is given by
\begin{equation}
w_i = \left\{
  \begin{array}{l l}
    4/9 & \quad i=0\\
    1/9 & \quad i=1\dots4\\
    1/36 & \quad i=5\dots8
  \end{array}  \right. \;\;
 \label{eq:appendix_wD2Q9}
\end{equation}
and
\begin{equation}
w_i = \left\{
  \begin{array}{l l}
    1/3 & \quad i=0\\
    1/18 & \quad i=1\dots6\\
    1/36 & \quad i=7\dots18
  \end{array}  \right. \;,
 \label{eq:appendix_wD3Q19}
\end{equation}
respectively.\\
The parameter $\phi^k_i$ is related to the compressibility of the fluid and, thus, to the speed of sound $c^k_s$ and hydrostatic pressure in phase $k$. For the D2Q9 and D3Q19 lattice $\phi^k_i$ is given by \cite{Leclaire2013:MultiPhase, Liu2012:RK3D}
\begin{equation}
\phi^k_i = \left\{
  \begin{array}{l l}
    \alpha_k & \quad i=0\\
    (1-\alpha_k)/5 & \quad i=1\dots4\\
    (1-\alpha_k)/20 & \quad i=5\dots8
  \end{array}  \right. \;\;
 \label{eq:appendix_phiD2Q9}
\end{equation}
and
\begin{equation}
\phi^k_i = \left\{
  \begin{array}{l l}
    \alpha_k & \quad i=0\\
    (1-\alpha_k)/12 & \quad i=1\dots6\\
    (1-\alpha_k)/24 & \quad i=7\dots18
  \end{array}  \right. \;,
 \label{eq:appendix_phiD3Q19}
\end{equation}
respectively, where one of the $\alpha_k$ is a free parameter setting the pressure level in the system.
The values of $\alpha_k$ are related by 
\begin{equation}
\gamma = \frac{\rho_\text{gas}}{\rho_\text{liq}} = \frac{1-\alpha_\text{liq}}{1-\alpha_\text{gas}}
 \label{eq:appendix_RelationAlpha}
\end{equation}
in order to guarantee a stable interface ($p_\text{gas}=p_\text{liq}$). In our simulations we set $\alpha_\text{gas}$ to 4/9.\\

\subsection{Two-phase collision: Perturbation}
The parameter $A$ controlling the surface tension is defined as \cite{Liu2012:RK3D}
  \begin{equation}
  A_\text{gas} = A_\text{liq} = A = \frac{9}{4} \sigma^\text{LBM} \bar{\tau} \;,
  \label{eq:appendix_SurfaceTension}
  \end{equation}
where $\sigma^\text{LBM}$ is the surface tension and $\bar{\tau}$ the density averaged relaxation time.
In our simulations we use a fourth-order isotropic approximation for the calculation of the color gradient \cite{Liu2012:RK3D}
\begin{equation}
\triangledown \psi = 3 \sum_{i=1}^{nv} w_i \; \mathbf{e}_i \; \psi(\mathbf{x}+\mathbf{e}_i)\;.
 \label{eq:appendix_ColorGradient3D}
\end{equation}
In the literature approximations of higher order are reported \cite{Sbragaglia2007}, however, the computational load increases significantly.\\
The parameters $B_i$ depend on the lattice and have to be chosen in order to ensure the conservation of mass in the perturbation step. For the D2Q9 and D3Q19 lattice the parameters are given by \cite{Reis2007:RK, Liu2012:RK3D}
\begin{equation}
B_i = \left\{
  \begin{array}{l l}
    -4/27 & \quad i=0\\
    2/27 & \quad i=1\dots4\\
    5/108 & \quad i=5\dots8
  \end{array}  \right. \;\;
 \label{eq:appendix_BD2Q9}
\end{equation}
and
\begin{equation}
B_i = \left\{
  \begin{array}{l l}
    -1/3 & \quad i=0\\
    1/18 & \quad i=1\dots6\\
    1/36 & \quad i=7\dots18
  \end{array}  \right. \;,
 \label{eq:appendix_BD3Q19}
\end{equation}
respectively.

\subsection{Two-phase collision: Recoloring}

The only additional parameter appearing in the recoloring operator is $\beta=0.85$ which determines the thickness of the inter-facial layer.

\end{document}